\documentclass[12pt,preprint,preprintnumbers,amsfonts,aps,nofootinbib]{article}

\usepackage{amssymb}
\usepackage{amsmath}
\usepackage{epsfig}

\usepackage{epsf}
\pdfoutput=1
\usepackage{graphicx}
\usepackage{amsfonts}
\usepackage{amssymb}
\usepackage{subfigure}

\usepackage{cite}

\def\mpl{M_{\rm Pl}}

\newcommand{\goodgap}{\hspace{\subfigtopskip}\hspace{\subfigbottomskip}}

\makeatletter
\renewcommand\section{\@startsection {section}{1}{\z@}%
                                 {-3.5ex \@plus -1ex \@minus -.2ex}
                                   {2.3ex \@plus.2ex}%
                                   {\normalfont\large\bfseries}}
\renewcommand\subsection{\@startsection{subsection}{2}{\z@}%
                                   {-3.25ex\@plus -1ex \@minus -.2ex}%
                                     {1.5ex \@plus .2ex}%
                                     {\normalfont\bfseries}}
\renewcommand\subsubsection{\@startsection{subsubsection}{3}{\z@}%
                                   {-3.25ex\@plus -1ex \@minus -.2ex}%
                                     {1.5ex \@plus .2ex}%
                                     {\normalfont\itshape}}
\makeatother

\def\beq{\begin{equation}}
\def\eeq{\end{equation}}
\def\be{\begin{equation}}
\def\ee{\end{equation}}
\def\bea{\begin{eqnarray}}
\def\eea{\end{eqnarray}}

\def\mpl{M_{pl}}

\def\e{\epsilon}

\def\d{\partial}

\DeclareRobustCommand{\SkipTocEntry}[4]{}

\begin{document}

\begin{titlepage}

\setcounter{page}{1} \baselineskip=15.5pt \thispagestyle{empty}

\begin{flushright}
\end{flushright}
\vfil

\begin{center}

{\Large \bf The Degeneracy Problem in Non-Canonical Inflation}
\\[0.9cm]
{\large Damien A. Easson$^{1,\dag}$ and Brian A. Powell$^{2,\star}$}
\\[0.5cm]
\vspace{.3cm}
{\normalsize {\sl  ${}^{1}$ Department of Physics  \& School of Earth and Space Exploration  \& Beyond Center\\
Arizona State University, Tempe, AZ 85287-1504}}
\\[0.2cm]
\vspace{.1cm}
{\normalsize {\sl ${}^2$ Institute for Defense Analyses, Alexandria, VA 22311}}\\

\vspace{.3cm}

\end{center}

\vspace{.8cm}

\hrule \vspace{0.3cm}
{\small  \noindent \textbf{Abstract} \\[0.3cm]
\noindent While attempting to connect inflationary theories to observational physics, a potential difficulty is the degeneracy problem: a single set of observables maps to a range of different inflaton potentials.  
Two important classes of models affected by the degeneracy problem are canonical and non-canonical models, the latter marked by the presence of a non-standard kinetic term that generates observables beyond the scalar and tensor two-point functions on CMB scales.  The degeneracy problem is manifest when these distinguishing observables go undetected.  We quantify the size of the resulting degeneracy in this case
by studying the most well-motivated non-canonical theory having Dirac-Born-Infeld Lagrangian. Beyond the scalar and
tensor two-point functions on CMB scales, we then consider the possible detection of equilateral non-Gaussianity at Planck-precision and a measurement of 
primordial gravitational waves from prospective space-based laser interferometers. The former detection breaks the degeneracy with canonical inflation but results in poor reconstruction prospects, while the latter measurement enables a determination of $n_T$ which, while not breaking the degeneracy, can be shown to greatly improve the non-canonical reconstruction. } \vspace{0.5cm}  \hrule

\vfil

\begin{flushleft}
{\normalsize { \sl \rm \small{${}^\dag$easson@asu.edu  \\ ${}^\star$brian.powell007@gmail.com}}}\\

\end{flushleft}

\end{titlepage}

\newpage
\tableofcontents
\newpage
\section{Introduction}
The most successful model of early universe cosmology is the inflationary universe paradigm \cite{Guth:1980zm,Linde:1981mu}. The general predictions
of single field slow roll inflation, including a large, flat universe with correlations between fluctuations in matter and radiation and a nearly scale-invariant spectrum of perturbations,  are all corroborated by the data; however, it is notoriously difficult to make any detailed statements about the fundamental mechanism behind the
accelerated expansion. In the simple case of single field canonical inflation, the amplitudes of the scalar and tensor power spectra on CMB scales uniquely map to the parameters       
of the inflationary Lagrangian. However, the inflaton need not be canonically normalized, nor need it be the primary source of primordial perturbations.  There might even be several inflaton fields cooperatively driving inflation.  In these cases, the unique mapping of observables to the Lagrangian is threatened -- distinct inflation models from different classes can give rise to the same set of observables, to within experimental precision.  This is known as the \it degeneracy problem\rm.   
Two broad classes of
theories which contribute to the degeneracy problem are theories in which degrees of freedom other than the inflaton serve as the source for primordial curvature
perturbations, \it e.g. \rm the curvaton scenario, and theories where the field dynamics are determined by terms in the Lagrangian other than, or in addition to, the scalar potential, $V(\phi)$, \it e.g. \rm models with non-canonical kinetic terms.  The former source of degeneracy -- that arising from curvatons -- was investigated in detail in a companion paper to the present work, Ref. \cite{Easson:2010uw}. In this paper, we examine the degeneracy problem in the context of non-canonical inflation, selecting the Dirac-Born-Infeld (DBI) model as a prototype.  

Exact degeneracies only exist between certain classes of models, for example, between canonical minimally and non-minimally coupled fields\footnote{This degeneracy does not exist between minimally and non-minimally coupled {\it non-canonical} models \cite{Easson:2009wc}.}.  Other broad classes, like curvaton and non-canonical models, generically give rise to distinguishing observables beyond the scalar and tensor power spectra, $P_\Phi(k)$ and $P_h(k)$, like isocurvature modes and/or non-Gaussianities.  Just as $P_\Phi(k)$ and $P_h(k)$ determine $V(\phi)$ in canonical inflation, in DBI an additional function, the sound speed $c_s$,  contributes to the reconstruction.  When $c_s \ll 1$, large equilateral non-Gaussianities are generated, and so $f^{\rm equil}_{NL}$ can be measured and used, along with $P_\Phi(k)$ and $P_h(k)$, to reconstruct $V(\phi)$ in DBI inflation.  In practice, the degeneracy problem is manifest when these additional observables are not detected, whether due to experimental limitations or the specifics of the scenario.  Planck will fail to detect $f^{\rm equil}_{NL}$ if $c_s \gtrsim 0.1$: then, not only would canonical inflation ($c_s=1$) be consistent with such an observation, but a range of other non-canonical model potentials that result from varying $c_s$ within the range $[0.1,1]$ would be equally consistent \cite{Powell:2008bi}.  The degeneracy problem is bad not only because we do not know the underlying type of theory ({\it e.g.} canonical or non-canonical), but because the resulting potential reconstruction is generally much worse relative to that possible under the assumption of canonical inflation for the same set of observables. 

In this paper, we perform Monte Carlo reconstructions of $V(\phi)$ for a range of observational outcomes.  We first consider the worst-case scenario in which no additional distinguishing observable, namely, equilateral non-Gaussianity, is detected.  We then examine how positive detections of $f^{\rm equil}_{NL}$ and a possible scale-dependence of the three-point function, $n_{NG}$, at Planck-precision affect the reconstruction.  While these observations break the degeneracy in favor of a model like DBI inflation, our reconstruction of $V(\phi)$ is limited by the precision with which these observables are measured.  Generally, we find that the DBI reconstruction is worse than the reconstructed canonical model generating the equivalent $P_\Phi(k)$ and $P_h(k)$.  That is, unless, it is possible to obtain a quality measurement of $n_T$, the tensor spectral index.  Such a measurement might be possible through a detection of primordial gravitational waves with future space-based laser interferometers such as the Big-Bang-Observer (BBO)
and Japan's planned DECIGO experiment.  In this case, even though the degeneracy is unbroken, the quality of the DBI reconstruction is greatly improved, with the range of consistent potentials only slightly larger than the range of canonical potentials.  This is because the modified consistency relation relating $n_T$ to the tensor/scalar ratio, $r=-8c_s n_T$, is used to replace the undetermined $c_s$ for $n_T$ in the reconstruction equations. This finding, described in detail in \S 3.4.3, was originally discussed along with a similar result applicable to curvatons in \cite{Easson:2010zy}.~\footnote{Note, the test of the inflationary consistency relations play a crucial role in helping to distinguish between different inflationary models, see e.g. \cite{Zhao:2011zb}.}

This paper is organized as follows. In \S 2, we outline the potential reconstruction program for non-canonical inflationary models. \S 3 provides the motivation for considering DBI models in the context of D-brane inflation in string theory and presents the Monte Carlo analysis.   \S 4 is devoted to a brief discussion of our results.

\section{Potential Reconstruction in Non-Canonical Inflation}
Methods for reconstructing a general inflationary Lagrangian were
developed in \cite{Bean:2008ga} and encompass a broad range of theories,
including non-canonical models in which the kinetic
term in the Lagrangian is not canonically normalized,
\be
S = \int d^4x \sqrt{-g}\left[\frac{\mpl^2}{2} R + \mathcal{L}(X,\phi)\right] 
\,,
\ee
where $X = \frac{1}{2}g^{\mu \nu}\d_\mu \phi \d_\nu \phi$.  Models of this form were first
popularized in the context of {\it k}-inflation \cite{ArmendarizPicon:1999rj}, in which
higher-derivative kinetic terms drive inflation in the absence of a potential.  More recently,
non-canonical inflation has found relevance in models constructed from string theory, such as D-brane
inflation \cite{Dvali:1998pa,Alexander:2001ks,Dvali:2001fw,Burgess:2001fx,Brodie:2003qv,Kachru:2003sx}
and the dynamical regime known as DBI inflation \cite{Silverstein:2003hf, Alishahiha:2004eh}.
While many of these constructions generically involve multiple fields, we restrict our analysis to
single field models.  

From the Friedmann and continuity equations,
\begin{eqnarray}
\label{Fnon}
\rho &=& 3M_{\rm Pl}^2H^2 = 2X\mathcal{L}_{X}-\mathcal{L},\\
\label{Fnon2}
\dot{\rho} &=& -3H(\rho+p) = -3H(\rho+\mathcal{L}),
\end{eqnarray}
one obtains the Hamilton-Jacobi equation \cite{Bean:2008ga},
\begin{equation}
\label{hjenc}
-\frac{\mathcal{L}(X,\phi)}{\mpl^2} = 3H^2(\phi) -\frac{4\mpl^2}{\mathcal{L}_X(X,\phi)}[H'(\phi)]^2,
\end{equation}
where $\mathcal{L}_X = \partial \mathcal{L}(X,\phi)/\partial X$.  Canonical inflation is
recovered for $\mathcal{L}_X = 1$, giving
\begin{equation}
\label{hje}
\frac{V(\phi)}{M_{\rm Pl}^2} = 3H(\phi)^2 - 2M_{\rm Pl}^2 [H'(\phi)]^2.
\end{equation}
Evidently, in canonical theories, $H(\phi)$ completely governs the inflationary trajectory and
determines the scalar potential $V(\phi)$.
In contrast, the non-canonical trajectory is characterized by two functions:
\begin{eqnarray}
\label{hje2}
H(\phi) &=& \sum_{n = 0}^{\infty} \frac{1}{n!}\frac{d^nH}{d\phi^n}(\phi - \phi_0)^n,\\
\label{hje3}
\mathcal{L}_X(\phi) &=& \sum_{n = 0}^{\infty} \frac{1}{n!}\frac{d^n\mathcal{L}_X}{d\phi^n}(\phi - \phi_0)^n,
\end{eqnarray}
resulting in two separate hierarchies of {\it flow parameters}.  The first, called the
{\it H-tower}, 
\begin{eqnarray}
\label{ncHflow}
H' &=& \frac{H}{M_{\rm Pl}}\sqrt{\frac{\e \mathcal{L}_X}{2}},\\
H'' &=& \frac{H\eta \mathcal{L}_X}{2M_{\rm Pl}^2},\nonumber \\
&\vdots& \nonumber \\
\frac{d^{(n+1)}H}{d\phi^{(n+1)}} &=& \left(\frac{\mathcal{L}_X}{2\mpl^2}\frac{H}{H'}\right)^n H'
\lambda_n,\nonumber\\
\end{eqnarray}
generalizes
the canonical flow parameters \cite{Liddle:1994dx}, recovered by taking $\mathcal{L}_X = 1$.
The second is unique to non-canonical theories and is called the { \it $\mathcal{L}_X$-tower},
\begin{eqnarray}
\label{ncLflow}
\mathcal{L}_X' &=& \frac{s}{2}\frac{H}{H'}\left(\frac{\mathcal{L}_X}{\mpl}\right)^2,\nonumber \\
\mathcal{L}_X'' &=& \frac{\varrho}{2}\left(\frac{\mathcal{L}_X}{\mpl}\right)^2,\nonumber \\
&\vdots& \nonumber \\
\frac{d^{(n+1)}\mathcal{L}_X}{d\phi^{(n+1)}} &=&
\left(\frac{\mathcal{L}_X}{2\mpl^2}\right)^n\left(\frac{H}{H'}\right)^{n-1} \mathcal{L}_X a_n.
\end{eqnarray}
With both $V(\phi)$ and the non-canonical kinetic term governing the inflationary
dynamics in these theories, cosmological observations must be used to disentangle the physical
effects of these two functions in order to successfully reconstruct them.  This is the challenge investigated in
this paper.

While many characteristics of non-canonical inflation are model dependent, 
we focus on the conservative models where
perturbations propagate relative to the homogeneous
background at speeds less
than that of light.~\footnote{For a more general treatment see~\cite{Babichev:2007dw}. }  
From Eqs. (\ref{Fnon}-\ref{Fnon2}), the hydrodynamical sound speed is given by
\begin{equation}
\label{cs}
c_s^2 = \frac{dp}{d\rho} = \frac{\mathcal{L}_{X}}{\mathcal{L}_{X} + 2X\mathcal{L}_{XX}} \leq 1,
\end{equation}
where $c^2 = 1$.
This is in contrast to canonical inflation, for which $\mathcal{L}_{XX} = 0$ and $c_s^2 = 1$.
Fluctuations that travel with speeds $c_s^2 < 1$ are non-Gaussian, a property that is not related
directly to the scalar potential; these observations will be important for distinguishing the
effects of $V(\phi)$ from those of the kinetic term.  In the next section, we  investigate
reconstruction in the context of a prototypical non-canonical model -- DBI inflation.

\section{Case Study: Reconstruction in DBI Inflation}

\subsection{The Scenario and the Power Spectrum}
A theoretically well-motivated example of non-canonical inflation arises from D-brane inflation in
string theory.  The most well-studied framework is based on Calabi-Yau flux compactifications of
Type IIB string theory \cite{Giddings:2001yu}, in which the inflaton parameterizes the motion of a probe brane
relative to an anti-brane at the tip of a warped throat geometry.  This geometry is given by the
line element \cite{Douglas:2006es},
\begin{equation}
ds^2_{10} = h^{-1/2}(y)g_{\mu \nu}{\rm d}x^\mu {\rm d}x^\nu + h^{1/2}(y)({\rm d}\rho^2 + \rho^2{\rm
d}^2_{X_5}),
\end{equation}
where the internal space is a cone over the base manifold $X_5$.  The throat coordinate $\rho$
determines the inflaton field, $\phi = \sqrt{T_3}\rho$, where $T_3$ is the tension of the brane: in
UV models the probe is a D3-brane falling towards a stack of anti-D3 branes at the tip of the throat,
while in IR models \cite{Chen:2005ad} the probe is an anti-D3-brane being pushed up the throat away from the stack.  In
either case, the equation of motion of the probe brane is obtained from the DBI Lagrangian \cite{Silverstein:2003hf},
\begin{equation}
\label{dbil}
\mathcal{L} = -f^{-1}(\phi)\sqrt{1+2f(\phi)X}+f^{-1}(\phi)-V(\phi),
\end{equation}
where $f(\phi)$ is determined by the warp factor $h^4(\phi)= T_3f(\phi)$.  The non-canonical
kinetic term in Eq. (\ref{dbil}) enforces an effective speed limit on the inflaton field, with the
result that `slow roll' inflation can be achieved even in the presence of steep potentials \cite{Alishahiha:2004eh}.
From the AdS/CFT correspondence, the speed limit on the field space arises from the fact that the probe brane must travel subluminally;
in analogy with special relativity, we define the $\gamma$-factor,
\begin{equation}
\label{gamma}
\gamma \equiv \frac{1}{\sqrt{1-f(\phi)\dot{\phi}^2}}.
\end{equation}
From the Lagrangian Eq. (\ref{dbil}), we find $\mathcal{L}_X = \gamma$ and using this in Eq. (\ref{cs})
gives $\gamma = c_s^{-1}$.  The causality constraint on the probe brane not only imposes a speed limit
on the field space, but also reduces the propagation speed of fluctuations.

The Friedmann equation follows from Eq. (\ref{dbil}),
\begin{equation}
3M_{\rm Pl}^2H^2(\phi)-V(\phi) = \frac{\gamma(\phi)-1}{f(\phi)},
\end{equation}
and gives
\begin{equation}
\dot \phi=-\frac{2M_{\rm Pl}^2}{\gamma(\phi)}H'(\phi).
\end{equation}
The $\gamma$-factor Eq. (\ref{gamma}) then becomes
\begin{equation}
\label{gammaH'}
\gamma=\sqrt{1+4M_{\rm Pl}^4f(\phi)\left[H'(\phi)\right]^2}.
\end{equation}
From these results the Hamilton-Jacobi equation, Eq. (\ref{hjenc}), can be obtained for DBI inflation,
\begin{equation} 
V(\phi)=3M_{\rm Pl}^2H^2(\phi)-4M_{\rm Pl}^4\frac{H'^2(\phi)}{\gamma(\phi)+1}.
\end{equation}
With $\mathcal{L}_X = \gamma$, we can use Eqs. (\ref{ncHflow}) and (\ref{ncLflow}) to write down the first few derivatives
of the potential, 
\begin{eqnarray}
\label{potdbi1}
V(\phi_{0})&=&M_{\rm Pl}^2H^2\left(3-2\epsilon\Gamma\right), \nonumber\\ V'(\phi_{0})&=&M_{\rm Pl}H^2\sqrt{2\epsilon\gamma}\left(3-2\eta\Gamma+s\Gamma^2\right),  \\ 
\label{potdbi3}
V''(\phi_{0})&=&
H^2\gamma\left[3\left(\epsilon+\eta\right)-2\left(\eta^2+\epsilon\lambda_2\right)\Gamma \right.
\nonumber \\  &&\left. +2s\left(2\eta-s\Gamma\right)\Gamma^2+\epsilon\varrho\Gamma^2\right],
\end{eqnarray}
where $\Gamma = {\gamma}/{(\gamma +1)}$.
An important distinction should be made between the DBI reconstruction and that of the curvaton carried out in the
companion analysis \cite{Easson:2010uw}: the curvaton does not modify the inflationary dynamics, it serves only
as a means of generating the primordial density perturbation.  In other words, the same inflationary Lagrangian
will generate different spectra depending on whether a curvaton is present or not -- a degeneracy with single
field models develops if we
are unable to detect its presence.  In contrast, DBI
inflation is dynamically distinct from canonical inflation, and so different observables necessarily arise from
different Lagrangians.  However, a degeneracy therefore develops if the observables characteristic of
non-canonical inflation are not detected.  

We proceed by connecting the
potential coefficients to cosmological observables.  In theories of inflation with an arbitrary speed
of sound, the amplitude of curvature perturbations at first order in slow roll is \cite{Garriga:1999vw}:
\begin{equation}
\label{scal}
P_{\Phi}(k) = \frac{1}{8\pi^2 M^2_{\rm Pl}}\left.\frac{H^2}{c_s \epsilon}\right|_{kc_s = aH},
\end{equation}
where the amplitude is evaluated at {\it sound} horizon-crossing: $kc_s = aH$.  On the other hand, the
tensor amplitude,
\begin{equation}
\label{tens}
P_h(k) = \frac{2}{\pi^2}\left.\frac{H^2}{M^2_{\rm Pl}}\right|_{k=aH},
\end{equation}
is evaluated at Hubble crossing, $k=aH$, because gravitational waves propagate at the speed of light.
If the difference in sound horizon crossing and Hubble crossing times is small for a given wavenumber,
$k$, then the tensor/scalar ratio is well approximated by the expression, 
\begin{equation}
\label{dbir}
r = 16c_s \epsilon.
\end{equation}
This is a fair assumption during slow roll \cite{Agarwal:2008ah,Powell:2008bi}.  
The scalar spectral index exhibits a dependence on the time rate
of change of $c_s$, 
\begin{equation}
\label{dbins}
n_s = 1 - 4\epsilon + 2\eta - 2s,
\end{equation}
while the tensor spectral index is the same as in canonical inflation,
\begin{equation}
n_T = -2\epsilon.
\end{equation}
Working to lowest order in slow roll and assuming a constant speed of sound ($s=0$), Eqs. (\ref{tens}),
(\ref{dbir}), and (\ref{dbins}) can be inverted to obtain the potential coefficients Eqs.
(\ref{potdbi1}-\ref{potdbi3}) in terms of the spectrum observables,
\begin{eqnarray}
\label{dbiPot1}
V(\phi_0) &=& \frac{\pi^2}{2} \mpl^4 P_\Phi r\left(3 - \Gamma 
\frac{r\gamma}{8}\right),\nonumber \\
V'(\phi_0) &=& \frac{\sqrt{2}\pi^2}{8} \mpl^3 P_\Phi r^{3/2}\gamma\left[3 - \Gamma\left(n_s-1 +
\frac{r\gamma}{4}\right)\right],\nonumber \\
\label{dbiPot3}
V''(\phi_0) &=& \frac{3\pi^2}{4} \mpl^2 P_\Phi r\gamma\left[n_s - 1 + \frac{3}{8}r\gamma \right.
\nonumber \\
&&\left.-\frac{8\Gamma}{3}\left(n_s-1+\frac{r\gamma}{4}\right)^2\right].
\end{eqnarray}
In additional to the spectrum observables, the potential depends on the value of the
$\gamma$-factor.\footnote{All expressions for DBI inflation reduce to those of canonical
inflation when $\Gamma = \frac{1}{2}$, $\gamma = 1$, $s = \varrho = \cdots = 0$.}
Despite the above-mentioned formal distinction between the curvaton and DBI reconstructions, the two cases
 are treated in a similar manor: determination of $V(\phi)$ requires observations of more than simply the adiabatic
density perturbation and tensor spectra.  In the case of curvatons, the amplitude of the curvaton fluctuation 
needs to be measured; in the case of DBI inflation, the $\gamma$-factor (equivalently, the sound speed
$c_s$) must be constrained.  Equations (\ref{dbiPot1}-\ref{dbiPot3}) are only
approximate and we will conduct a more rigorous reconstruction in a later section.
\subsection{Non-Gaussianity}
The non-canonical kinetic term in the DBI Lagrangian Eq. (\ref{dbil}) generically leads to the production of
primordial non-Gaussianity.  In contrast to curvaton and general multifield models, the large non-Gaussianity
arises from the three-point correlator of the field fluctuations, $\langle \delta \phi_{{\bf k}_1} \delta \phi_{{\bf
k}_2} \delta \phi_{{\bf k}_3}\rangle$, rather than the nonlinear interaction of the field
and curvature perturbations.   In the DBI model, the strongest
correlation arises amongst modes of comparable wavenumber at the horizon, and so $f^{(3)}_{NL}$ is evaluated in the {\it equilateral}
limit, ${\bf k}_1 = {\bf k}_2 = {\bf k}_3$ \cite{Alishahiha:2004eh,Chen:2006nt},
\begin{equation}
\label{dbifnl}
f^{(3)}_{NL} = f^{\rm equil}_{NL} = -\frac{35}{108}\left(\gamma^2 - 1\right).
\end{equation}
Being of a different form in Fourier space than local-type non-Gaussianity (for which ${\bf k}_1 \approx {\bf k}_2 \gg
{\bf k}_3$), a measurement of the shape of non-Gaussianities offers the possibility of distinguishing DBI
inflation from the curvaton and other theories that generate local non-Gaussianities, like modulated reheating.  The
non-Gaussianities produced in DBI inflation can be very large in the DBI limit, $\gamma \gg 1$, in which
the field is rolling relativistically, $\dot{\phi}^2 \simeq f^{-1}(\phi)$.  The equilateral-type non-Gaussianities are
less constrained by current data than the local-type: WMAP7 gives $-214 < f^{\rm equil}_{NL} < 266$ at 95\% CL \cite{Komatsu:2010fb}.
This is due to the fact that for a given amplitude, the distribution of local-type fluctuations is more significantly
skewed than the equilateral distribution.  As a result, future projections are also not as limiting as for local
non-Gaussianities: Planck should achieve $|\Delta f^{\rm equil}_{NL}| \sim
26$ at 68\% CL \cite{Baumann:2008aq}.  However, in contrast to the
non-Gaussianities produced in the curvaton scenario, a measurement of $f^{\rm equil}_{NL}$ completely determines the additional
degree of freedom in the reconstruction, and a detection of higher-order statistics like the trispectrum is
unnecessary.  

For completeness we mention that it is possible that the amplitude of non-Gaussianities generated during DBI inflation varies with scale
\cite{Chen:2005fe}, so that
$f_{NL} \sim k^{n_{NG}}$, with the spectral index defined as
\begin{equation}
\label{nngdbi}
n_{NG} = \frac{d\ln f_{NL}}{d\ln k}.
\end{equation}
The projected 1$\sigma$ error on $n_{NG}$ expected from Planck is \cite{Sefusatti:2009xu}
\begin{equation}
\label{errornngdbi}
\Delta n_{NG} \simeq 0.3\frac{100}{f^{\rm equil}_{NL}},
\end{equation}
with CMBPol expected to improve on this limit by around a factor of two. We consider Planck sensitivity detections
for non-Gaussianity in this
paper as the factor of two improvement will not be significant
enough to strongly affect the results.  We investigate the effect of a detection
of non-Gaussianities -- both the amplitude $f_{NL}^{\rm equil}$ and the scale dependence $n_{NG}$ -- on
reconstruction in \S \ref{dbinong}.

\subsection{Gravitational Waves}
The non-standard sound speed that is the hallmark of DBI inflation results in the modified scalar spectrum given by Eq.
(\ref{scal}).  However, since gravitational waves propagate at the speed of light, the spectrum of tensor modes is
still proportional to the energy scale of inflation, Eq. (\ref{tens}).
This results in a modified consistency
relation,
\begin{equation}
\label{dbicon}
r = -8c_s n_T.
\end{equation}
A reliable determination of the consistency relation requires a quality measurement of
$n_T$.
Concept studies of future space-based laser interferometers designed to measure the primordial gravitational wave
signal on small scales ($\sim 0.1-1$ Hz) have yielded promising results.  Two proposals,
the Big Bang Observer (BBO) \cite{bbo} and Deci-hertz Interferometer Gravitational Wave
Observatory, (DECIGO) \cite{Seto:2001qf}, will detect gravitational waves if B modes on CMB scales give $r \gtrsim 10^{-3}$ and $r \gtrsim           
10^{-6}$, respectively -- a substantial part of the observable parameter space.  These
probes are intended to enable a determination of $n_T$ with a precision surpassing
future CMB probes: BBO should give a 65\% CL of $\Delta n_T \sim 10^{-2}$ while DECIGO
might achieve $\Delta n_T \sim 10^{-3}$ or better \cite{Seto:2005qy,Kudoh:2005as}.
We
simulate the effect that such a detection has on DBI reconstruction in
the next section.
\subsection{Monte Carlo Analysis}
The inversion from the observable parameter space $(r, n_s, dn_s/d\ln k,\cdots)$ to
the flow space $(\epsilon, \eta, \gamma, \lambda_2,\cdots)$ enables one to write the inflaton
potential in terms of observables as in Eq. (\ref{dbiPot1}).  However, this process is
only tractable if the observables are written to lowest order in the flow parameters
(c.f. Eqs. (\ref{dbir}) and (\ref{dbins})).  A more accurate and efficient method for
reconstruction at higher order can be accomplished using Monte Carlo methods
\cite{Hoffman:2000ue,Kinney:2002qn,Easther:2002rw}.  One begins with the flow parameters
$\epsilon, \eta, \lambda_2, \gamma, s, \cdots$ and for each stochastically sampled initial
condition, solves the flow equations \cite{Peiris:2007gz},
\begin{eqnarray}
\label{eq:flowequations}
\epsilon &=& \frac{1}{H}\frac{d H}{d N},\cr
\frac{d \epsilon}{d N} &=& \epsilon\left(2 \eta - 2 \epsilon - s\right),\cr
\frac{d \eta}{d N} &=& -\eta\left(\epsilon + s\right) +  \lambda_2,\cr &\vdots& \cr
\frac{d \lambda_\ell}{d N} &=& -\lambda_\ell \left[\ell \left(s + \epsilon\right) - \left(\ell - 1\right) \eta\right]
+ \lambda_{\ell +1},\cr
s &=& \frac{1}{\gamma}\frac{d \gamma}{d N},\cr
\frac{d s}{d N} &=& -s \left(2 s + \epsilon - \eta\right) + \epsilon \varrho,\cr
\frac{d \varrho}{d N} &=& -2 \varrho s + a_2,\cr &\vdots& \cr
\frac{d a_\ell}{d N} &=& -a_\ell \left[\left(\ell + 1\right) s + \left(\ell - 1\right) (\epsilon-\eta)\right] +
a_{\ell+1}, \cr
\frac{d\ln k}{dN} &=& \epsilon + s -1,
\end{eqnarray}
where the system is truncated by taking $\lambda_{M+1} = \alpha_{M+1} = 0$ and $d N = -H dt$ is the number of efolds before the end of inflation.  Taken as a
time variable, $\Delta N < 0$ over the course of inflation.  
The solution to these equations is an exact DBI inflationary trajectory with potential
determined from Eq. (\ref{dbiPot1}).  The observables can also be determined from the
flow parameters and those models which agree with observations can be selected out.
We retain only those models that support sufficient inflation to generate the CMB
anisotropies, $|\Delta N| \gtrsim 10$, and we work to $6^{th}$-order in
the $H$-tower and $3^{rd}$-order in the $\gamma$-tower.  We draw the $H$-tower
parameters from the initial ranges
\begin{eqnarray}
\label{ics}
\epsilon_i &\in& [0,0.8],\nonumber \\
\lambda_{\ell,i} &\in& 10^{-\ell + 1}[-0.5,0.5],
\end{eqnarray}
and the $\gamma$-tower from the ranges: $\gamma \in [1,30]$, $s\in [-0.1,0]$, $\varrho \in [-0.01,0.01]$.  We are therefore including in our analysis models
with a time-dependent speed of sound; from Eq. (\ref{hje3}) with $\mathcal{L}_X = \gamma$, the inverse sound speed is quadratic in $\phi$.  From Eq.
(\ref{gammaH'}), there is then some freedom in the form of the warp factor, $h^4(\phi) = T_3f(\phi)$.     

We choose $\ln k_i = -8.047$ so that the largest length scales of interest correspond to
the quadrupole. 
Since density perturbations freeze out when $kc_s = aH$ and tensors when $k=aH$, care must be taken when matching the scale
$k_{obs}$ to the e-fold number, $N_{obs}$.  For density perturbations, the scale at
horizon crossing is determined by solving $d{\rm ln}k/dN = \epsilon + s -1$ along with the flow equations, and for each acceptable solution we calculate the observables at $k = 0.01\,{\rm Mpc}^{-1}$
\cite{Chen:2006nt,Kinney:2007ag}
\begin{eqnarray}
P_{\Phi} &=& \left[1-2\epsilon-2s + 2b\left(\epsilon-\frac{\eta}{2}+\frac{s}{2}\right)\right]\frac{1}{8\pi^2M_{\rm Pl}^2}\left.\frac{H^2}{c_s \epsilon}\right|_{kc_s= a H}, \nonumber \\
n_{s}&=& 1-4\epsilon+2\eta-2s-2(1+C)\epsilon^2-(3+C)s^2 \nonumber \\ 
&&-\frac{1}{2}(3-5C)\epsilon\eta-\frac{1}{2}(11+3C)\epsilon s+(1+C)\eta s \nonumber \\
&& +\frac{1}{2}(1+C)\epsilon \varrho +\frac{1}{2}(3-C) (\lambda_2), \nonumber \\ 
\alpha &=& -\left(\frac{1}{1-\epsilon-s}\right)\frac{dn_{s}}{dN}.
\end{eqnarray}
where $b = 2 - {\rm ln}2 - \gamma$, and $\gamma = 0.5772$ is the Euler-Mascheroni constant.  For tensor perturbations, the scale at horizon crossing is given by the usual relation $d{\rm ln}k/dN = 1 - \epsilon$, with the following observables
calculated at $k = 0.01\,{\rm Mpc}^{-1}$ \cite{Lorenz:2008et},
\begin{eqnarray}
P_h &=& \left[1-2(1-b)\epsilon\right]\left.\frac{2H^2}{\pi^2\mpl^2}\right|_{k=aH},\nonumber \\
n_T &=& -2\epsilon -2\epsilon s + (4b-6)\epsilon^2 - 4(b-1)\epsilon \eta.
\end{eqnarray}
We then calculate the tensor/scalar ratio from $r = P_h(k_0)/P_\Phi(k_0)$.  

We perform the above-described flow analysis for canonical inflation as well, obtained
with
the above method by taking $\gamma = 1$, $s = \varrho =\cdots = 0$.  The approach taken here is
analogous to that carried out in \cite{Easson:2010uw} for curvatons: the results
obtained there for single field field inflation are identical to those obtained here for
canonical inflation, and the results that we present here for DBI models can also be directly
compared against our findings for curvatons in \cite{Easson:2010uw}.  
We now present results for different
observational outcomes.  
\subsubsection{No Detection of Non-Gaussianity}
\begin{figure*}
\centering
$\begin{array}{ccc}
\subfigure[]{
\includegraphics[width=0.32 \textwidth,clip]{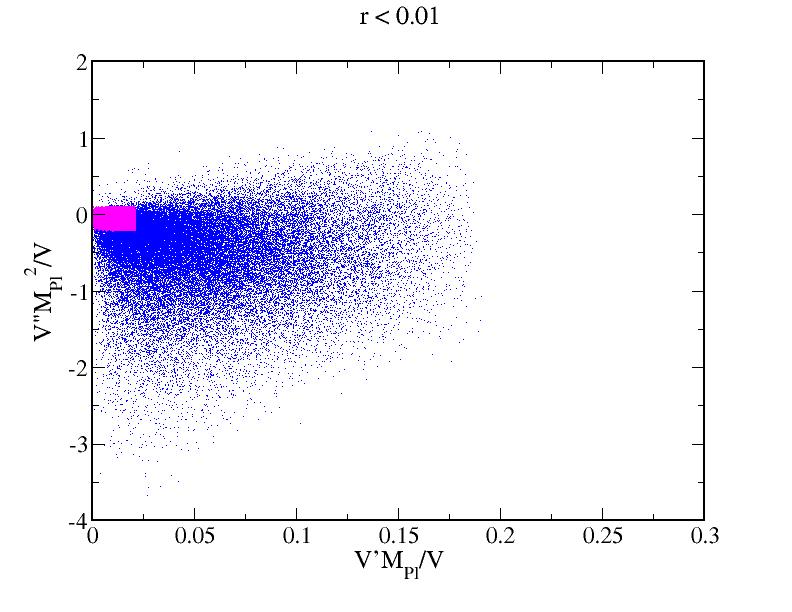}}
\subfigure[]{
\includegraphics[width=0.32 \textwidth,clip]{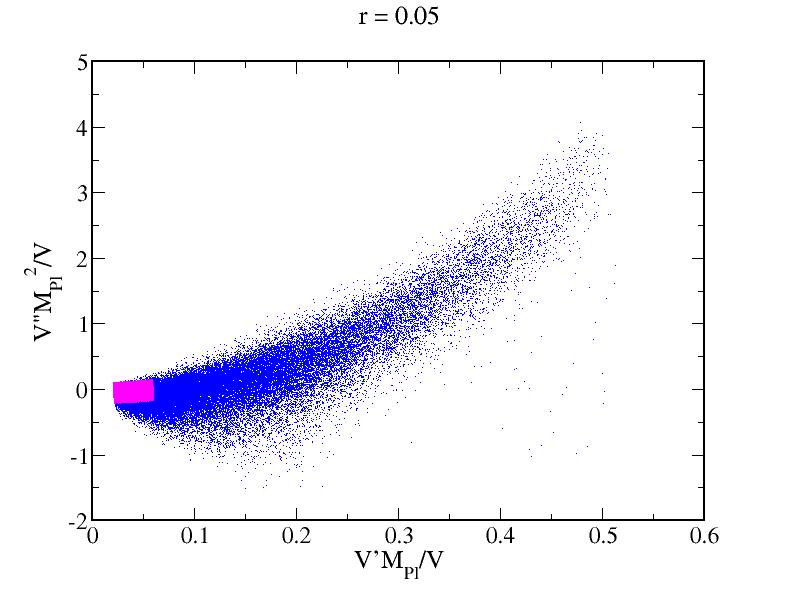}}
\subfigure[]{
\includegraphics[width=0.32 \textwidth,clip]{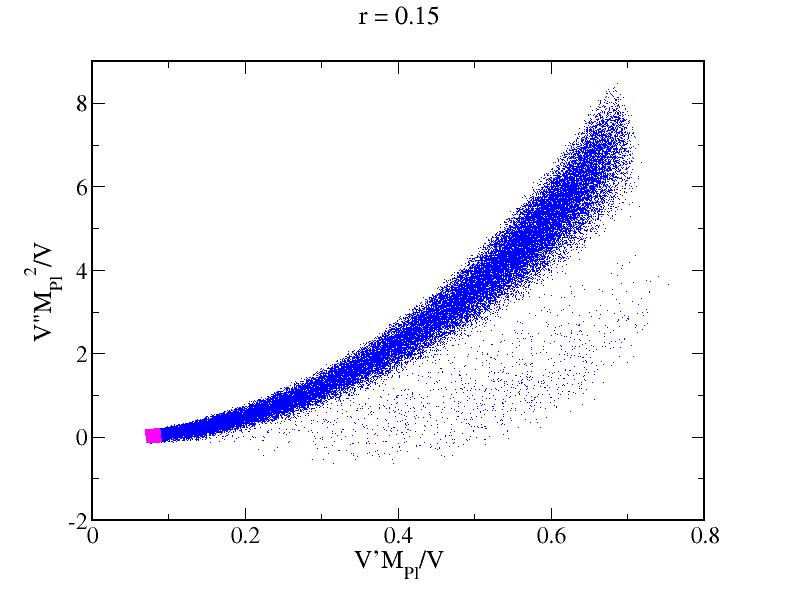}}\\
\end{array}$
$\begin{array}{ccc}
\subfigure[]{
\includegraphics[width=0.32 \textwidth,clip]{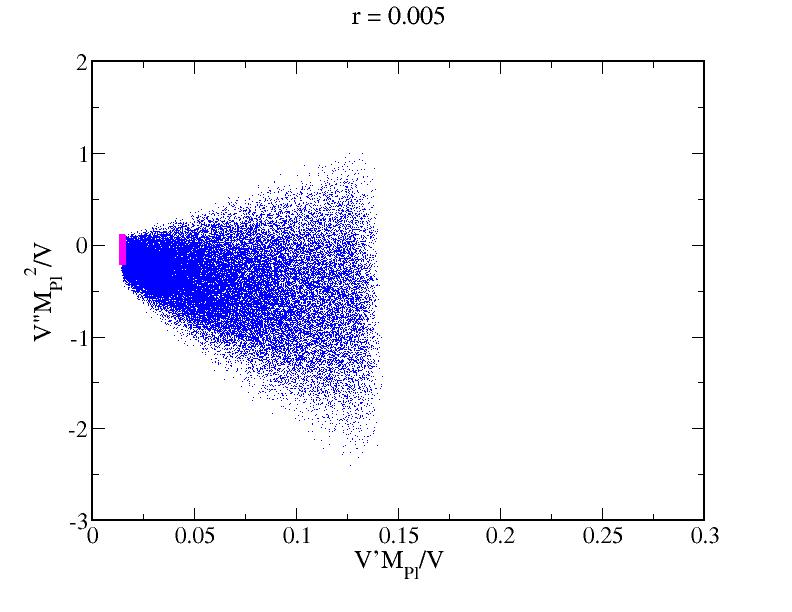}}
\subfigure[]{
\includegraphics[width=0.32 \textwidth,clip]{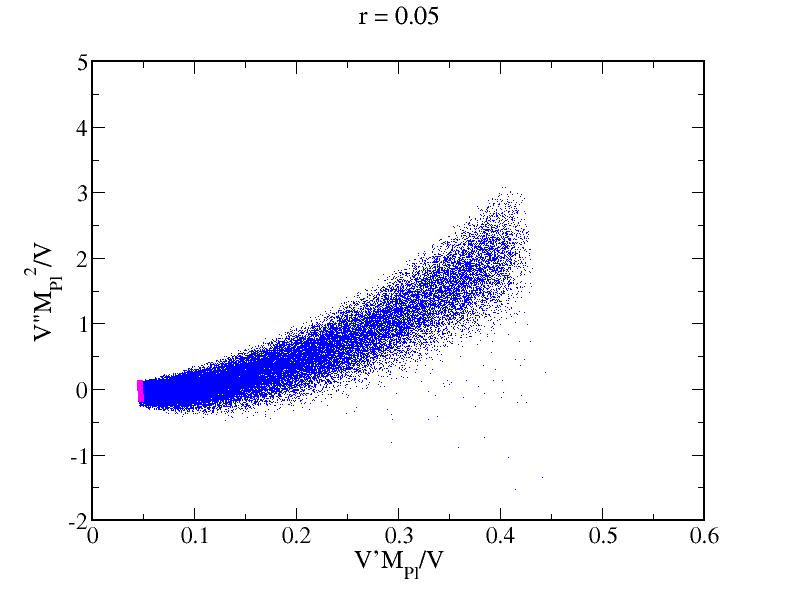}}
\subfigure[]{
\includegraphics[width=0.32 \textwidth,clip]{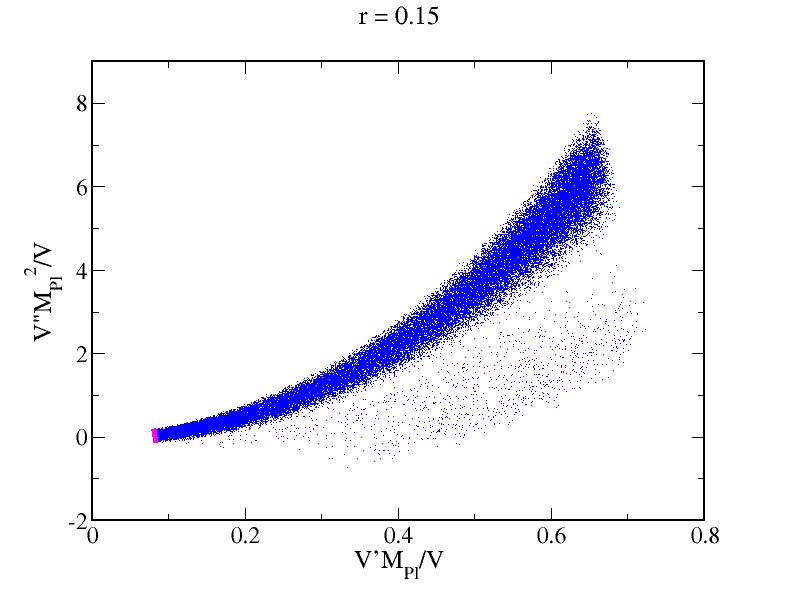}}
\end{array}$
\caption{Monte Carlo results of single field (magenta) and DBI (blue) reconstructions making use of only the spectral observables
$r$, $n_s$, and $dn_s/d{\rm ln}k$.  We
present results for three fiducial values of $r$: $r = 0.005$ (a),(d), $r = 0.05$ (b),(e), $r = 0.15$ (c),(f).  Top row presents results for Planck and bottom row for CMBPol.}
\end{figure*}
In this section we perform Monte Carlo reconstruction of DBI inflation at Planck-precision in the event that
non-Gaussianities are not detected, corresponding to $f^{\rm equil}_{NL} \geq -26$ ($\gamma \lesssim
9$) (68\% CL) \cite{Baumann:2008aq}.  This case was first
investigated using both Bayesian and flow methods in \cite{Powell:2008bi}\footnote{In Ref. \cite{Powell:2008bi}, a more optimistic detection threshold for $f^{\rm equil}_{NL}$ than that considered here was used.}.  
We consider projected constraints from Planck and CBMPol: for Planck we consider 68\% CL detections of $r$ ($r \gtrsim 0.01$, $\Delta r \sim 0.03$) \cite{Colombo:2008ta}, $n_s$ ($\Delta
n_s \sim 0.0038$), and $dn_s/d{\rm ln}k$ ($\Delta dn_s/d{\rm ln}k \sim 0.005$)
\cite{Bond:2004rt}, and for CMBPol we assume 68\% CL detections of $r$ ($r \gtrsim 10^{-4}$, $\Delta r \sim r/10$), $n_s$ ($\Delta n_s \sim 0.0016$), and $dn_s/d{\rm ln}k$
($\Delta dn_s/d{\rm ln}k \sim 0.0036$) \cite{Baumann:2008aq}.
We again consider three different fiducial values for the tensor/scalar ratio: $r=0.005$, $r=0.05$ and $r=0.15$,
and fix $n_s = 0.97$ and $dn_s/d\ln k = 0$.  In Figure 1 we present 50,000 models each of canonical inflation (magenta) and DBI inflation (blue).  There is a significant disparity between the constraints on $V(\phi)$ in the
case of DBI versus canonical inflation, consistent with the findings of \cite{Powell:2008bi}.  The uncertainties in
$V'/V$ and $V''/V$ are about a factor of 10 and 12 greater, respectively, for DBI
relative to canonical inflation with fiducial $r=0.05$.  For $r=0.15$, the uncertainties are increased by more than an order
of magnitude.  Similar to our findings in \cite{Easson:2010uw}, the reconstruction achievable with Planck in the
absence of tensors ($r \lesssim 0.05$) is comparable to that achieved with a detection of tensors by CMBPol (with fiducial $r= 0.005$).
The fact that CMBPol fails to improve constraints on $V(\phi)$ in the presence of both a curvaton and DBI
degeneracy suggests that this problem is widespread amongst degenerate models.  It must be stressed that a
determination of $r$ still serves to reliably constrain the energy scale of inflation, even amongst models
described by more general Lagrangians.  And, we will see in the case of DBI inflation, as we did with curvatons in
\cite{Easson:2010uw}, that a measurement of both $r$ and $n_T$ will prove to be quite advantageous to the
reconstruction effort \cite{Easson:2010zy}.  However, the results in Figure 1 indicate that improvements in a measurement of $r$ on its
own offer little corresponding improvement in reconstruction results. 

Before investigating how additional observables can improve on the degenerate case, we examine the effect of
the degeneracy on the zoology classification of inflation models  \cite{Dodelson:1997hr,Kinney:2003uw}.  In
canonical single field inflation, models fall into distinct families with unique observable predictions in the
$(n_s,r)$ plane.  Hybrid models are generally monotonic polynomial potentials with nonvanishing vacuum energy at their minimum
satisfying $V''(\phi) > 0$ and $({\rm log}V(\phi))'' > 0$.  Large field models are also generally monotonic polynomials but
with a true vacuum at the minimum.  In order to obtain sufficient inflation with these kinds of potentials, the
initial field displacement must be large in Planck units.  Large field models satisfy $V''(\phi) > 0$ and $({\rm
log}V(\phi))'' < 0$.  Lastly, small field models are monotonic polynomials only near their local maxima -- they
prototypically resemble the potentials of `new inflation' and those governing spontaneous symmetry breaking.  They
satisfy $V''(\phi) < 0$ and $({\rm log}V(\phi))'' < 0$.  We present the zoology for canonical single field
inflation in Figure 2 (a).  At lowest order, the regions are delineated by the observable conditions,
\begin{eqnarray}
r &=& -8(n_s-1) \,\,\,\, (\rm{large \,\, field - hybrid}),\\
r &=& -\frac{8}{3}(n_s-1) \,\,\,\, (\rm{small \,\, field - large \,\, field}).
\end{eqnarray}
While a zoology exists for DBI inflation, the conditions that define boundaries of the observable regions depend
on $\gamma$,
\begin{eqnarray}
r &=& -\frac{8}{\gamma}(n_s-1) \,\,\,\, (\rm{large \,\, field - hybrid}),\\
r &=& -\frac{8}{3\gamma}(n_s-1) \,\,\,\, (\rm{small \,\, field - large \,\, field}).
\end{eqnarray}
It is apparent that any uncertainty in $\gamma$ will translate into an uncertainty in exactly where to draw these
delineations -- there will exist regions in which models cannot be uniquely classified according to the zoology.
These regions are degenerate in that they result from the overlap of at least two classes.  In the case just
considered in which non-Gaussianities are not discovered in future missions, we find an uncertainty $\Delta \gamma
\approx 9$, and the resulting zoology shown in Figure 2 (b). 
\begin{figure*}
\centering
$\begin{array}{cc}
\subfigure[]{
\includegraphics[width=80mm]{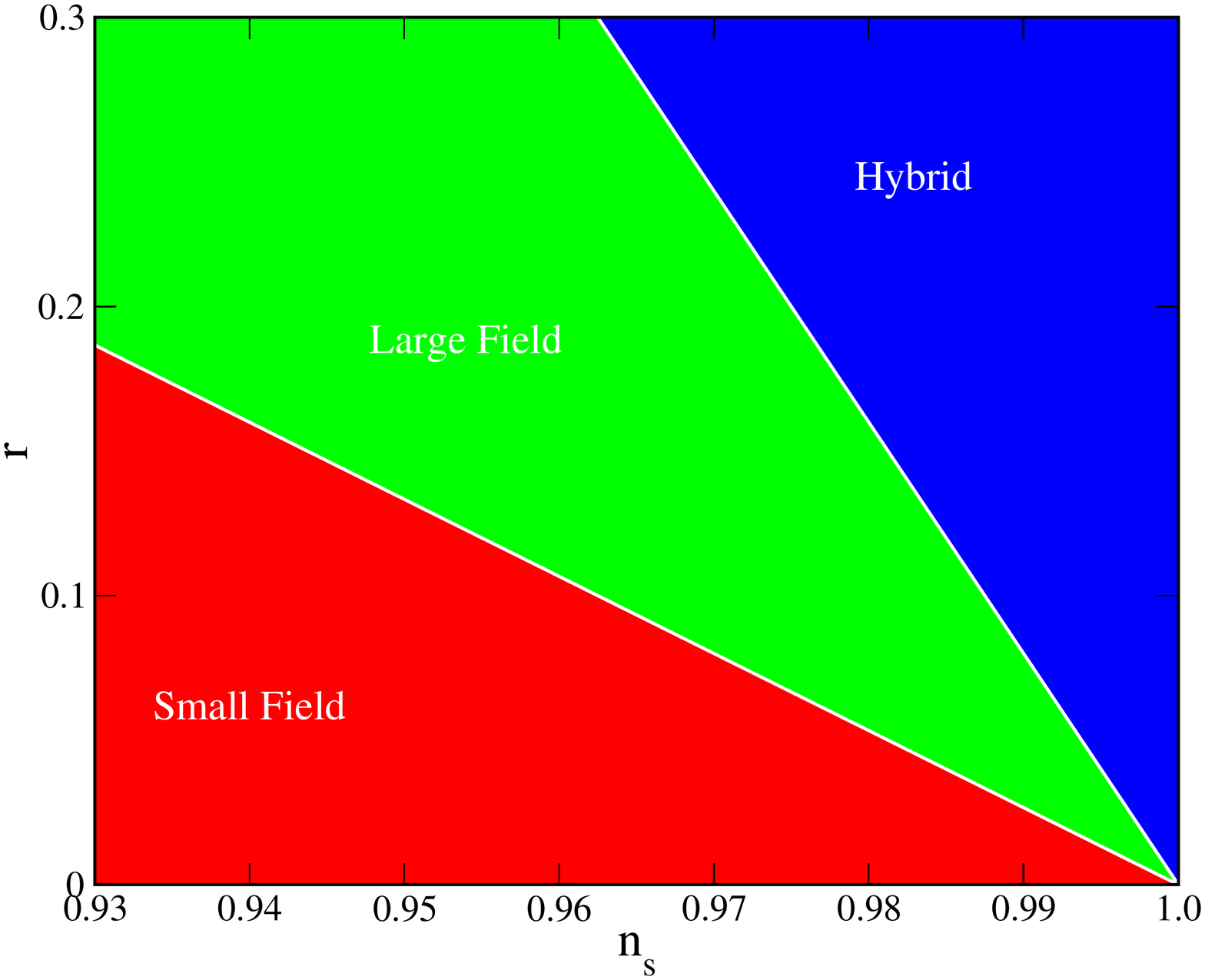}}
\subfigure[]{
\includegraphics[width=80mm]{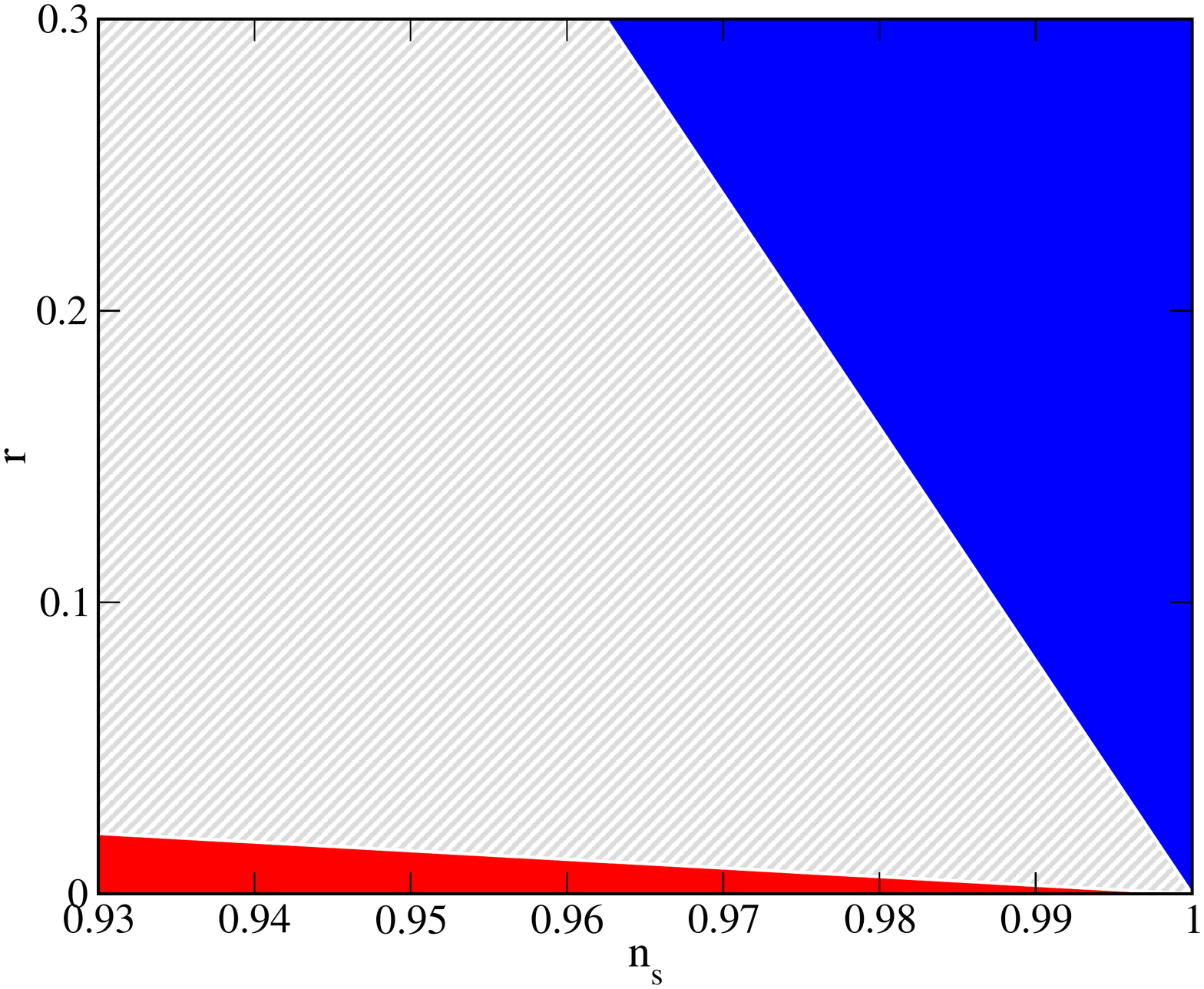}}
\end{array}$
\caption{(a) Canonical single field inflationary zoology.  (b) Zoology in the presence of the DBI degeneracy.
Models that fall within the gray region cannot be uniquely assigned to a class.}
\end{figure*}
The gray region in Figure 2 (b) corresponds to a large degenerate area: hybrid DBI models overlap large
field canonical models, and large field DBI models overlap a large portion of the small field canonical region.

\subsubsection{Detection of Non-Gaussianity}
\label{dbinong}
\begin{figure*}[htp]
\label{dbi1}
\centering
$\begin{array}{cc}
\subfigure[]{                                                                     
\includegraphics[width=80mm]{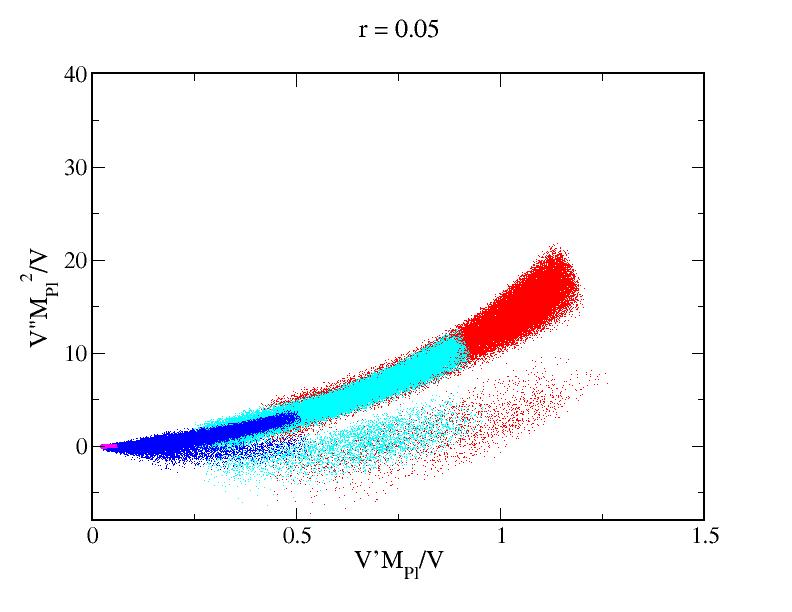}}                                          
\subfigure[]{                                                                     
\includegraphics[width=80mm]{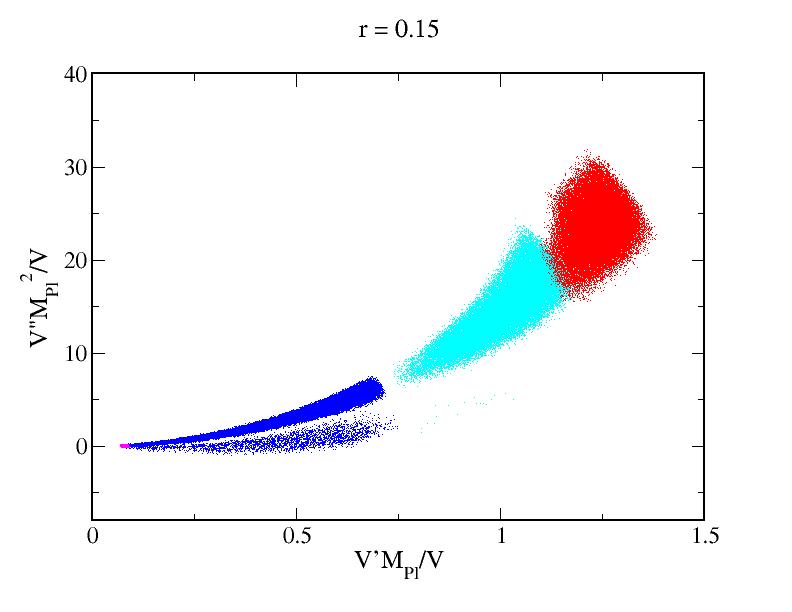}}
\end{array}$
\caption{Monte Carlo results of DBI reconstruction for two possible detections of $f^{\rm equil}_{NL}$: $-70 \pm 26$ (cyan)
and $-150 \pm 26$ (red).  The canonical (magenta) and DBI (blue) reconstructions in the absence of
non-Gaussianities from Figure 1 are included for context.  We
present results for two fiducial values of $r$: (a) $r = 0.05$ and (b) $r = 0.15$.}
\end{figure*}
We next consider the case in which equilateral non-Gaussianities are detected in the future at Planck-precision.  While such an observation breaks any degeneracy with canonical inflation, the reconstruction is still limited by the precision with which the degeneracy-breaking observables are measured.  Throughout this analysis, we therefore include the reconstructed canonical models generating the equivalent $P_\Phi(k)$ and $P_h(k)$ for comparison. 

For each fiducial value of the tensor/scalar
ratio we present results in Figure 3 for two possible
observations: $f^{\rm equil}_{NL} = -70 \pm 26$ (cyan) and $f^{\rm equil}_{NL} = -150 \pm 26$ (red).  We retain the models from Figure 2 for
context.  The Planck-sized error on $f^{\rm equil}_{NL}$ translates to a 68\% CL uncertainty in the
$\gamma$-factor of $\gamma \in [11,17]$ for $f^{\rm equil}_{NL} = -70$ and $\gamma \in
[19,23]$ for $f^{\rm equil}_{NL} = -150$.  As the value of $\gamma$ increases, the resulting
central values of $V'/V$ and $V''/V$ shift, although the shift is only greater than
the individual errors in the potential parameters for moderate values of $r$, as seen by comparing
Figure 3 (a) and
(b). Meanwhile, since $f^{\rm equil}_{NL} \propto \gamma^2$, the error on $\gamma$ is
reduced for larger non-Gaussianities and we consequently expect the uncertainties in
$V(\phi)$ to be smaller as $\gamma$ is increased, although this is only evident in Figure 3 (b) for $V'/V$.  The
unexpected increase in the error on $V(\phi)$ between the non-detection
of non-Gaussianities (blue) and the positive detection (cyan and red), seen
particularly in Figure 3 (a), is instead a result of
the fact that the we have taken the 68\% limit for a null detection of non-Gaussianities to
coincide with the 68\% CL associated with a positive detection\footnote{While not
strictly correct, in comparison with the order of magnitude disparity between the
canonical and DBI reconstructions, the distinction between the 68\% limit of the hypothesis
test and the 68\% CL of the detection is unimportant.}.  This, in
conjunction with the constraint $f^{\rm equil}_{NL} < 0$, results in a one-tailed
distribution for the null case and a smaller error.   We therefore see that in no case does a
a measurement of the amplitude  $f^{\rm equil}_{NL}$ result in a clear improvement in both $V'/V$ and
$V''/V$ over a null detection.

In addition to a detection of the amplitude $f_{NL}^{\rm equil}$, we also examined the effects that a measurement of the running Eq.
(\ref{nngdbi}) has on reconstruction.   From Eq. (\ref{dbifnl}), and with 
\begin{equation}
\frac{dN}{d\ln k} = \frac{1}{\epsilon +s-1},
\end{equation}
where $s$ is defined as in Eq. \ref{ncLflow}, we find
\begin{equation}
n_{NG} = \frac{2s}{\epsilon + s - 1}.
\end{equation}
We see that $\gamma \sim c_s^{-1}$ is unconstrained by a measurement of $n_{NG}$, and instead the higher-order parameter, $s \propto
\dot{c_s}$, largely determines the scale dependence.   This fact, coupled with the large projected errors on
$n_{NG}$ Eq. (\ref{errornngdbi}), results in no improvement in constraints on $V(\phi)$ with $n_{NG}$ over the
degenerate case.

\subsubsection{Direct Detection of Primordial Gravitational Waves}
\begin{figure*}[htp]
\label{MC}
\goodgap \goodgap \goodgap \goodgap \goodgap \goodgap \goodgap \goodgap \goodgap                                                                                  
\goodgap \goodgap \goodgap \goodgap \goodgap 
\subfigure[]{
\includegraphics[width=0.32 \textwidth,clip]{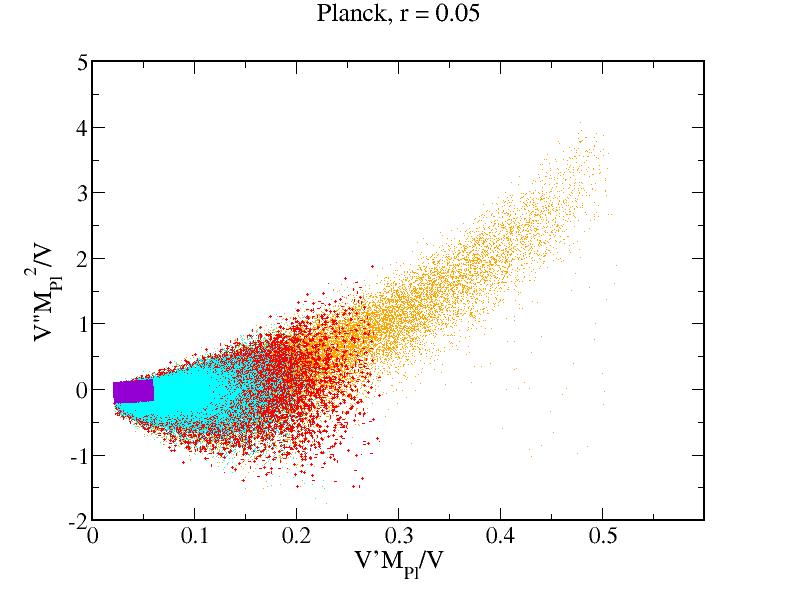}}
\subfigure[]{
\includegraphics[width=0.32 \textwidth,clip]{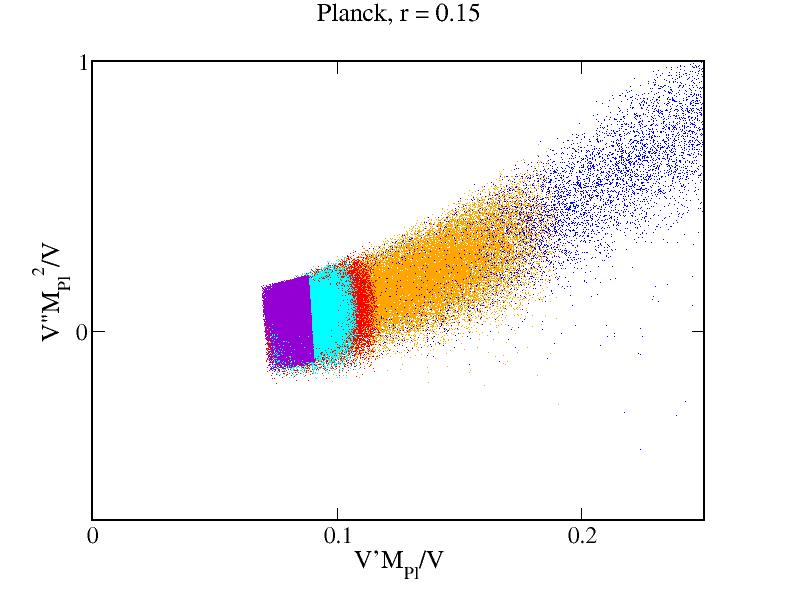}}
$\begin{array}{ccc}
\subfigure[]{
\includegraphics[width=0.32 \textwidth,clip]{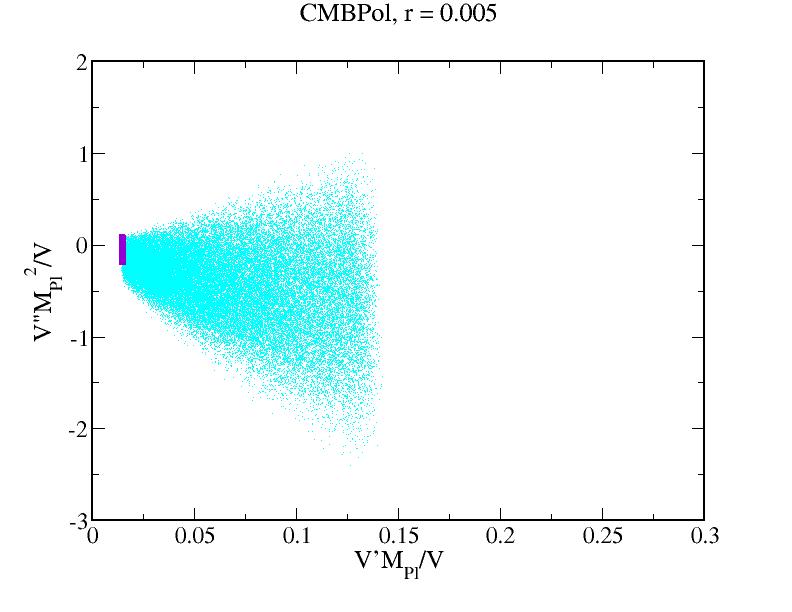}}
\subfigure[]{
\includegraphics[width=0.32 \textwidth,clip]{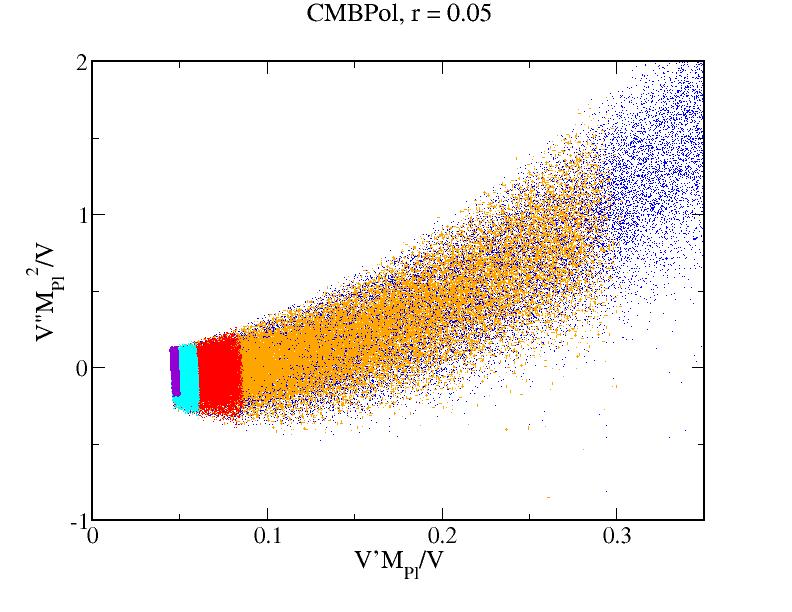}}
\subfigure[]{
\includegraphics[width=0.32 \textwidth,clip]{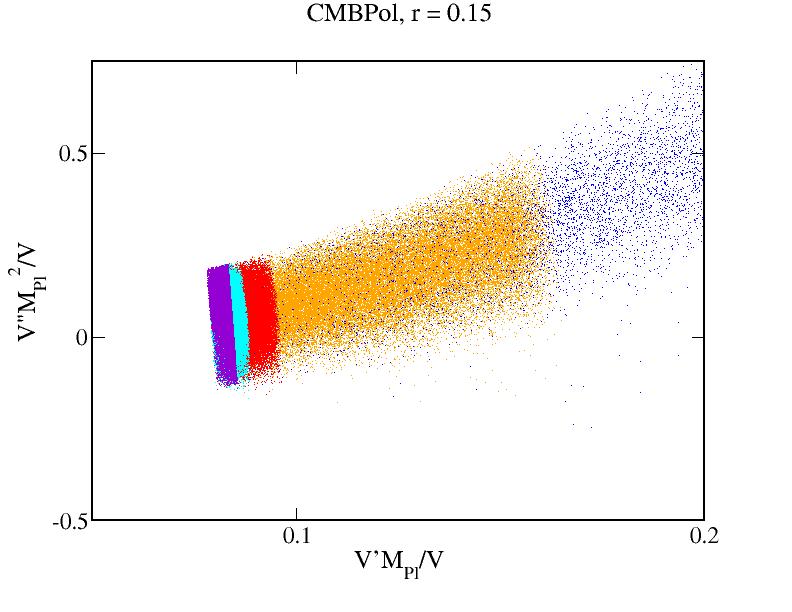}}
\end{array}$
$\begin{array}{ccc}
\subfigure[]{
\includegraphics[width=0.32 \textwidth,clip]{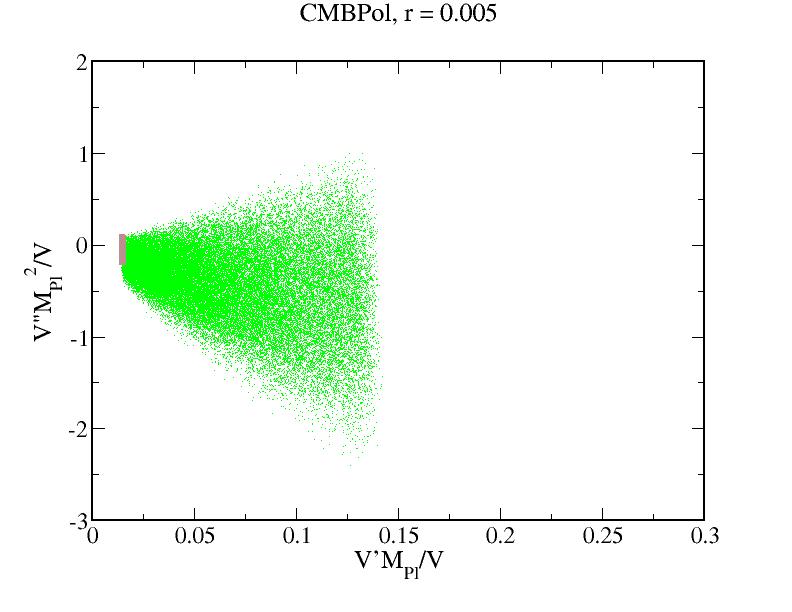}}
\subfigure[]{
\includegraphics[width=0.32 \textwidth,clip]{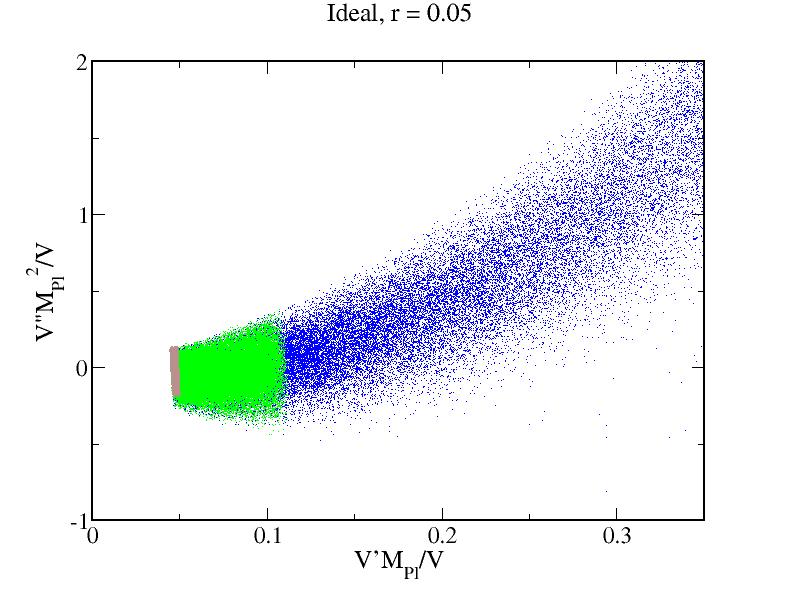}}
\subfigure[]{
\includegraphics[width=0.32 \textwidth,clip]{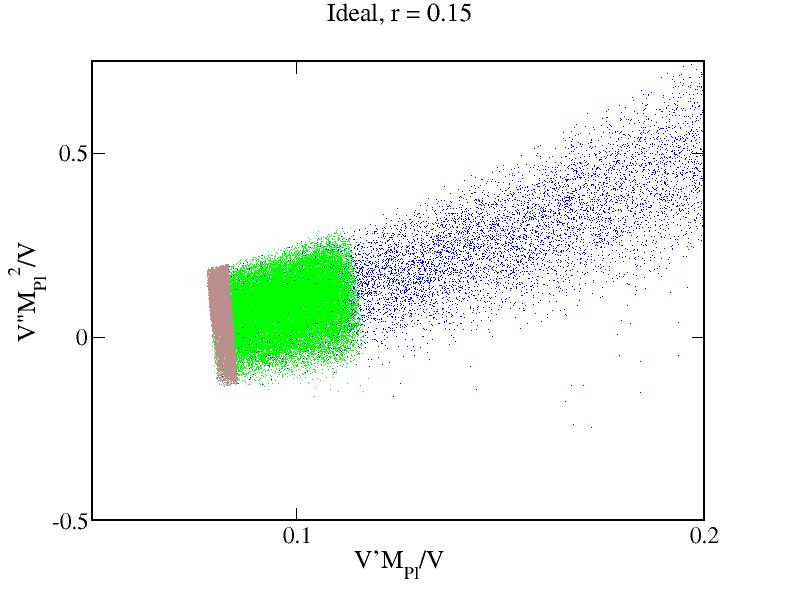}}
\end{array}$
\caption{Monte Carlo results for canonical and DBI models with a direct detection of tensors
at three different precisions: BBO-standard, BBO-grand, and DECIGO. We present models
with three different fiducial values: $(r = 0.005; n_T = -0.000625)$, $(r = 0.05; n_T = -0.0625)$,
and $(r = 0.15; n_T = -0.01875)$. DBI models are orange (BBO-standard), red (BBO-grand), cyan (DECIGO), and blue
points give the reconstruction with CMB data alone (same points as Fig. 2 for the respective satellite missions.) In (a)
we note that the orange points overlap the blue points, indicating that BBO-standard offers no improvement over Planck alone.
For single field models we present DECIGO only in purple. We consider
different precisions on the measurement of $r$: Planck (top row) and CMBPol (middle row).
In the bottom row we present forecasts for an ideal B mode detection: DBI models in green and single field models in gray. 
The blue points in the bottom row show the reconstruction with CMBPol alone.}

\end{figure*}
As a last case, we study the effect that a direct detection of primordial gravitational
waves has on DBI reconstruction.  We consider possible detections with three prospective probes:
BBO-standard, BBO-grand, and DECIGO.  
This analysis provides additional details in support of results presented in \cite{Easson:2010zy}. We assume a tensor spectrum of the form
\begin{equation}
P_h(k) = P_h(k_0)\left(\frac{k}{k_0}\right)^{n_T + \frac{1}{2}\alpha_T{\rm ln}\left(\frac{k}{k_0}\right)},
\end{equation}
where $\alpha_T = dn_T/d{\rm ln}k$ is the tensor index running, and determine the uncertainty in the measurement of $n_T$ from
\begin{equation}
\label{errornt}
\Delta n_T = \left\{\left[\frac{6\times 10^{-18}}{XA_{GW}P_h(k_*)}\right]^2 + \left[\frac{1}{2}\alpha_T{\rm ln}\left(\frac{k_*}{k_0}\right)\right]^2\right\}^{1/2},
\end{equation}
where $\alpha_T \simeq 4\epsilon \eta - 4\epsilon^2 - 2\epsilon s $ and $A_{GW} = 2.74
\times 10^{-6}$, and $X$ characterizes the experiment: $X = 0.25$, 2.5, 100 for
BBO-standard, BBO-grand, and DECIGO, respectively.  
The amplitude of the gravitational wave signal, $\Omega_{GW}$, is constrained on the scale of direct detection ($k_* = 6.5 \times
10^{14}$ Mpc$^{-1}$) to lie within the 68\% CL, $\Delta \Omega_{GW} = X\Omega_{GW}/10^{-18}$.
We choose fiducial values $(r,n_T,\Omega_{GW})$ consistent with a
power law spectrum $\Omega_{GW}h^2 = A_{GW}rP_\Phi(k_0)(k_*/k_0)^{n_T(k_*)}$ that obey the canonical single field consistency
relation $n_T(k_*) = -8/r(k_0)$.

This choice of fiducial values is in agreement with both canonical and DBI inflation, and so unlike the non-Gaussianities just
considered, primordial gravitational waves do not break the degeneracy in this case.
However, a precision measurement of $n_T$, coupled with a determination of $r$, nonetheless enables an improved reconstruction of
$V(\phi)$ despite the persistent degeneracy because of the modified consistency relation, Eq.
(\ref{dbicon}).  The potential coefficients, Eqs. (\ref{dbiPot1}), become
\begin{eqnarray}
V'(\phi_0) &=& -\frac{V_0}{M_{\rm Pl}}n_T \sqrt{\frac{8}{r}},\\
V''(\phi_0) &=& -\frac{4V_0}{M^2_{\rm Pl}r}\left(n_s - 1 - 3n_T\right),
\end{eqnarray}
with no longer any dependency on $\gamma$.  
We present results in Figure 4 for the fiducial
values: $(r=0.005$, $n_T = -0.000625)$, $(r = 0.05$, $n_T = -0.00625)$, and $(r=0.15$, $n_t = -0.01875)$.  We find that for
$r=0.005$, direct detection offers no improvement over the degenerate case shown in Figure 1 (d); however, improvements begin to emerge as the value of $r$ increases.
With a detection of $r \gtrsim 0.05$ with CMBPol and $n_T$ with DECIGO, the DBI
reconstruction becomes comparable to that of canonical inflation (Figures 4 (d),(e)).  Even the less
optimistic outcome of a Planck detection of tensors with $r \gtrsim 0.05$ and a
BBO-grand measurement of $n_T$, we find that the DBI potential space can be constrained
to within a factor of a few of the canonical reconstruction (Figures 4 (a),(b)).  

For completeness and comparison, we also consider the reconstruction that follows from
the determination of $n_T$ from an ideal B mode measurement on CMB scales (Figures 4
(f)-(h)).  
We take the uncertainties in $n_T$ for the ideal experiment to be $\Delta n_T = 0.01$ at $r=0.005$, $\Delta n_T = 0.009$ at $r=0.05$, and $\Delta
n_T = 0.007$ at $r=0.15$ \cite{Zhao:2009mj}.
The constraints are slightly worse but competitive with those obtained with 
CMBPol+BBO-grand.  We summarize all of our findings in Table 1.  

Given the improved constraints on $V(\phi)$ obtainable with $n_T$, we now revisit the zoology
classification.  Earlier, we obtained the zoology in the absence of $n_T$, Figure 2 (b),
and found that much of the available $(n_s,r)$ parameter space could not be uniquely
assigned to a potential class.  This was a result of the fact that $\gamma$ was
virtually unconstrained.  However, with a detection of $n_T$, in many cases the limits
of $\gamma$ are much improved (c.f. Table 1): $\Delta \gamma \sim \mathcal{O}(1)$ for
$r \sim 0.15$ with Planck and $r \gtrsim 0.05$ with CMBPol and an ideal experiment.
Rather than determine the zoology for each combination in Table 1, we present in Figures
5 (a)-(c) the zoology that results for $\Delta \gamma = 1$, 0.5, and 0.1.  The zoology
is almost restored for $\Delta \gamma = 0.01$, obtained with CMBPol+BBO-grand or DECIGO
for $r \gtrsim 0.15$.
\begin{figure*}
\centering
$\begin{array}{ccc}
\subfigure[]{
\includegraphics[width=0.33 \textwidth,clip]{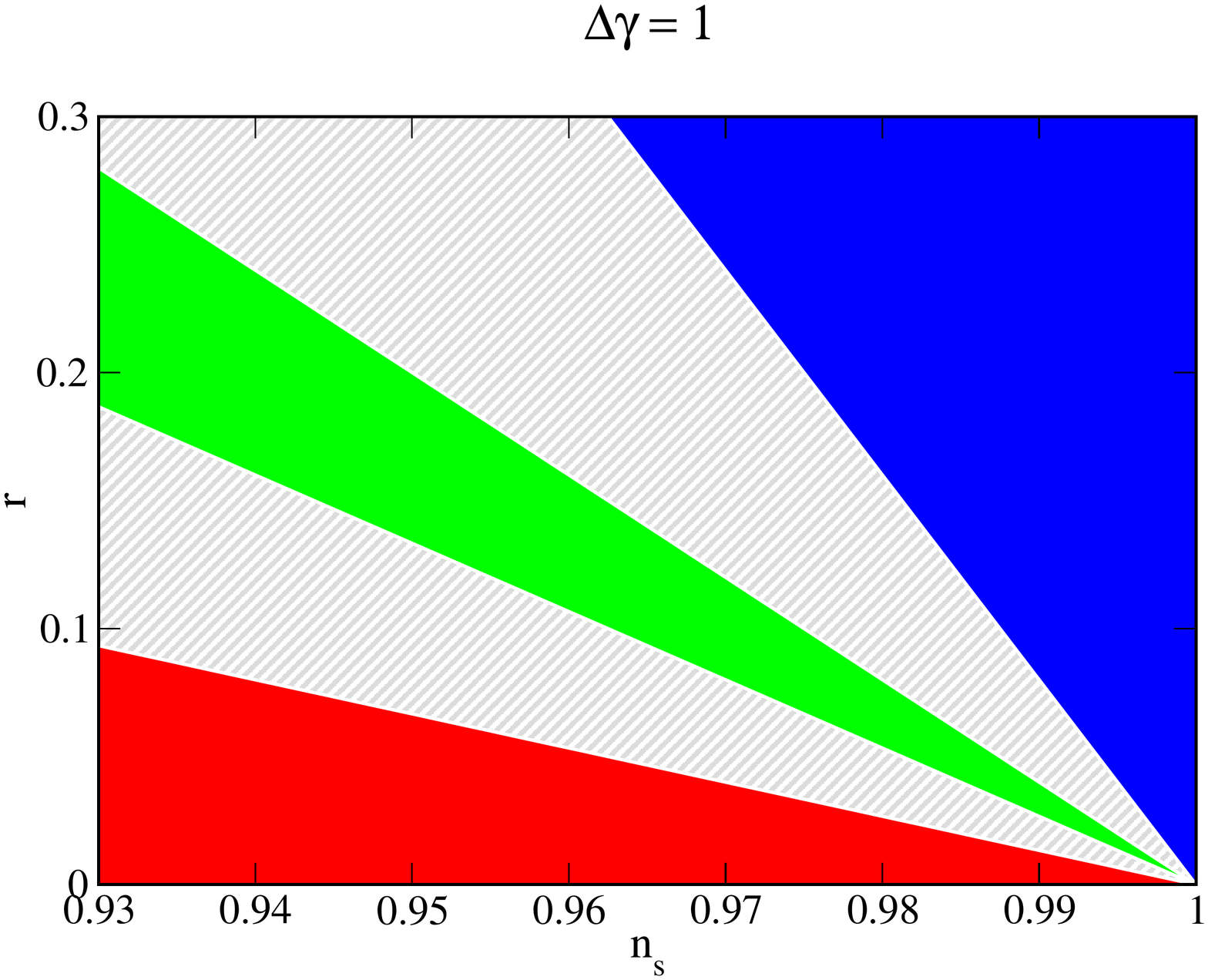}}
\subfigure[]{
\includegraphics[width=0.33 \textwidth,clip]{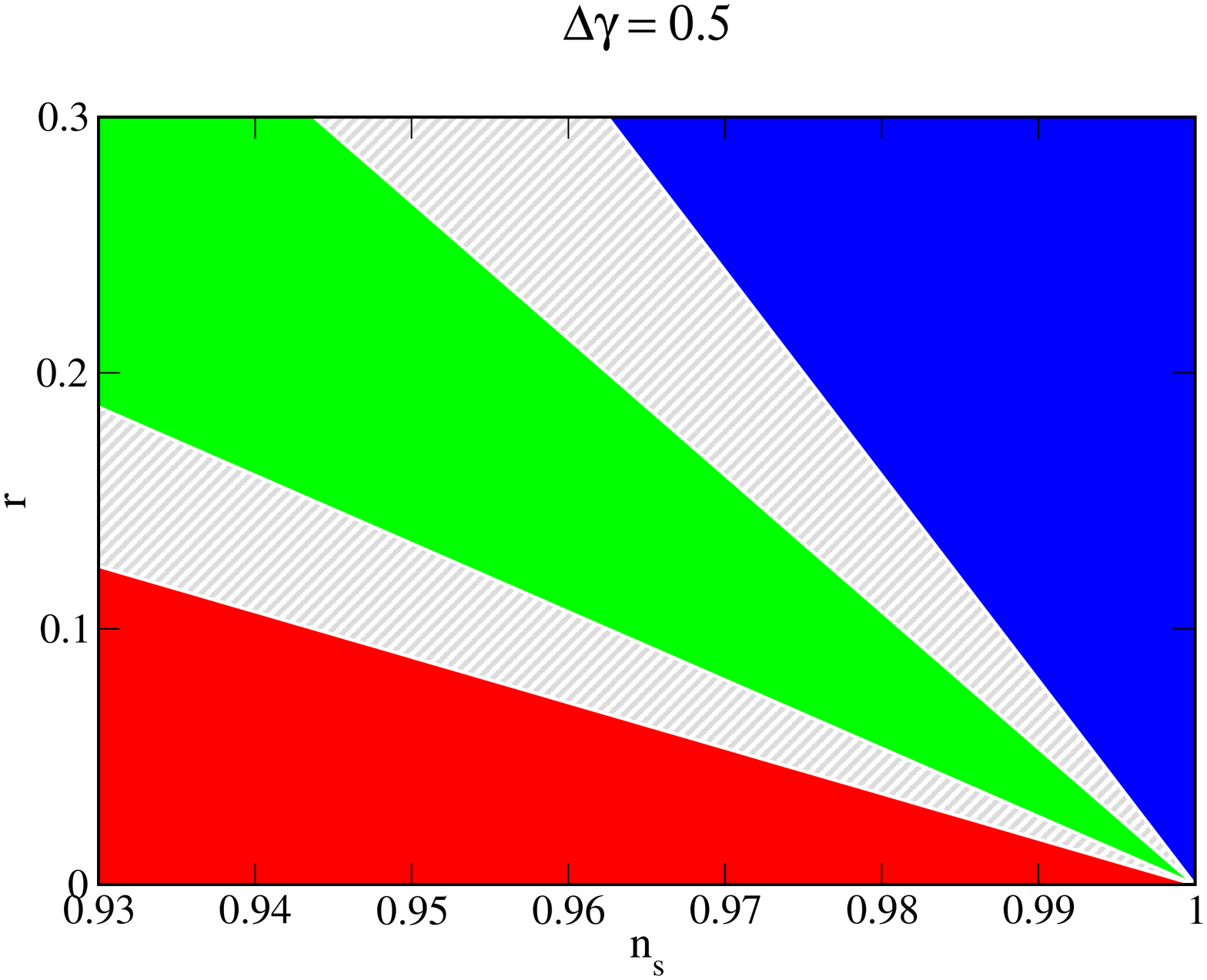}}
\subfigure[]{
\includegraphics[width=0.33 \textwidth,clip]{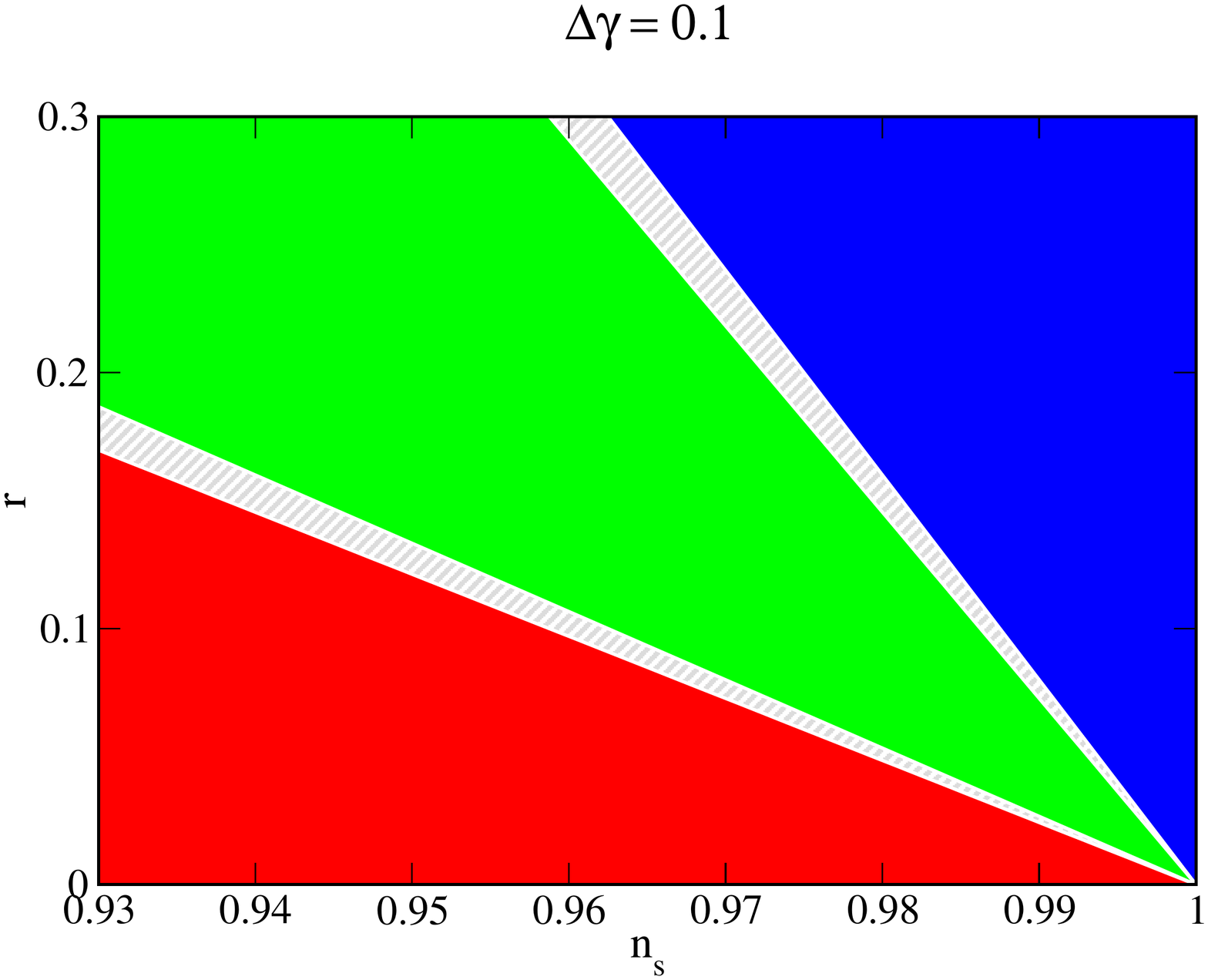}}\\
\end{array}$
\caption{Zoology in the presence of a DBI degeneracy where $\gamma$ has been constrained by varying degrees
with, for example, a measurement of $n_T$.  We present three cases: (a) $\Delta \gamma =1$, (b) $\Delta \gamma =
0.5$, and (c) $\Delta \gamma = 0.1$.}
\end{figure*}
\begin{table*}
\tiny
\begin{tabular}{|l|cc|cc|c|c|}
\hline
\hline
Observation & $\Delta V'_{\rm DBI}/\Delta V'_{\rm single}$ &$\Delta V''_{\rm DBI}/\Delta V''_{\rm single}$ &$\Delta
V'_{\rm DBI}/\Delta V'_{\rm deg}$&$\Delta V''_{\rm DBI}/\Delta V''_{\rm deg}$ &$\Delta \gamma$ & Zoology unique?\\
\hline
No NG/$n_T$ &9/28 &12/14 &1&1&9 & H \\
NG ($f^{\rm equil}_{NL} = -70$)&18/84&24/42&2/3&2/3 &5.5 & H \\
NG ($f^{\rm equil}_{NL} = -150$)&9/14&24/28&1/0.5&2/2 &3 & H \\
NG ($n_{NG}$)&9/28 &12/14&1&1 &9 & H \\
$n_T$ (Planck + BBOs) &9/7 &12/2&1/0.25&1/0.0625& 400/2 & H \\ 
$n_T$ (Planck + BBOg) &5/2.5 &8/1.25&0.5/0.08&0.5/0.03& 55/0.75&H \\ 
$n_T$ (Planck + DEC) &3/2 &5/1.25&0.25/0.05&0.25/0.03& 40/0.5& H/LF,H \\
$n_T$ (CMBPol + BBOs) &30/40/15 &11/6/2&1/0.6/0.2&1/0.25/0.0625& 125/15/1 & H/H/H\\ 
$n_T$ (CMBPol + BBOg) &30/8/3 &11/1.5/1&1/0.1/0.03&1/0.08/0.03 & 125/1/0.15 & H/H/LF,H\\ 
$n_T$ (CMBPol + DEC) &30/3/1.75 &11/1.25/1&1/0.035/0.015&1/0.08/0.03 & 40/0.5/0.1 & H/SF,H/LF,H\\
$n_T$ (ideal B mode) &30/16/7 &11/2/1 &1/0.2/0.1 &1/0.1/0.05 &125/1.75/0.5& H,H,LF,H \\
\hline
\end{tabular}
\caption{Reconstruction results of each case considered in this analysis.  In the first
set of columns the errors on the potential coefficients for DBI inflation are given relative to those expected
from single field inflation; in cases where the observation rules out single field inflation ({\it e.g.} large
non-Gaussianities), the values are given for
reference only.  In the second set of columns these errors are given relative to the
worst-case degeneracy (no measurement of non-Gaussianity or $n_T$.)  When constraints depend on the fiducial value of $r$ chosen, we provide results separated by a
slash, with the constraints for $r=0.005$ followed by those for $r=0.05$ and $r=0.15$ (for Planck only these latter two values are relevant);  if only one number is
provided then there is no difference.  In the last column, `All' indicates that all three zoology classes
can be uniquely reconstructed; otherwise, only those classes that can be uniquely reconstructed are listed.  NG: non-Gaussianities, H:
hybrid, LF: large field, SF: small field.}
\normalsize
\end{table*}
%

\section{Discussion}
In this paper we have studied the degeneracy problem in the context of non-canonical inflation, focusing on DBI inflation as a prototype. Our main goal in this endeavor has been to provide an estimate of the size of degeneracy that exists between canonically normalized single field inflationary models and DBI models in the event that no observables beyond the scalar and tensor two-point functions are measured. This `worst case scenario' makes it impossible to reconstruct the inflationary potential or categorize the inflationary model according to the standard zoology scheme. This is of course only the `tip of the iceberg', as many other $k$-inflation models and (even other paradigms such as the curvaton as discussed in \cite{Easson:2010uw})  exacerbate the problem. We then demonstrate the ability of additional observables such as non-Gaussianity, B-mode polarization and direct detection of primordial gravitational waves to both break the degeneracy and improve the potential reconstruction program.

Our results show that even though detection of the amplitude of $f^{\rm equil}_{NL}$ clearly breaks the degeneracy between models the detection does not improve reconstruction of the inflationary potential. Then we show that direct measurement of the tensor perturbation spectral index $n_T$, which does not help to break degeneracy, does result in significantly improved potential reconstruction when combined with an accurate determination of the tensor-to-scalar ratio $r$ via the modified consistency relation \cite{Easson:2010zy}. Without the measurement of $n_T$,
measurements of $r$, even at precision greater than that achievable by Planck, does not lead to an improvement in potential reconstruction.  In the best case scenario, measurements of $r$ and $n_T$ (which serves to constrain the sound speed in DBI inflationary models) can result in an impressive restoration of the inflationary zoology. 
%

\section*{Acknowledgments}
The work of DAE is supported in part by 
the DOE under DE-SC0008016 and by the Cosmology Initiative at Arizona State University.


\begin{thebibliography}{999}
\bibitem{Guth:1980zm} 
  A.~H.~Guth,
  Phys.\ Rev.\ D {\bf 23}, 347 (1981).

\bibitem{Linde:1981mu} 
  A.~D.~Linde,
  Phys.\ Lett.\ B {\bf 108}, 389 (1982).
  
\bibitem{Easson:2010zy}
  D.~A.~Easson and B.~A.~Powell,
  Phys.\ Rev.\ Lett.\  {\bf 106}, 191302 (2011)
  [arXiv:1009.3741 [astro-ph.CO]].

\bibitem{Easson:2010uw}
  D.~A.~Easson and B.~A.~Powell,
  Phys.\ Rev.\  D {\bf 83}, 043502 (2011)
  [arXiv:1011.0434 [astro-ph.CO]].

 \bibitem{Easson:2009wc} 
  D.~A.~Easson, S.~Mukohyama and B.~A.~Powell,
  Phys.\ Rev.\ D {\bf 81}, 023512 (2010)
  [arXiv:0910.1353 [astro-ph.CO]]. 
\bibitem{Bean:2008ga}
  R.~Bean, D.~J.~H.~Chung and G.~Geshnizjani,
  Phys.\ Rev.\  D {\bf 78}, 023517 (2008)
  [arXiv:0801.0742 [astro-ph]].
\bibitem{ArmendarizPicon:1999rj}
  C.~Armendariz-Picon, T.~Damour and V.~F.~Mukhanov,
  Phys.\ Lett.\  B {\bf 458}, 209 (1999)
  [arXiv:hep-th/9904075].
\bibitem{Zhao:2011zb} 
  W.~Zhao and Q.~-G.~Huang,
  Class.\ Quant.\ Grav.\  {\bf 28}, 235003 (2011)
  [arXiv:1101.3163 [astro-ph.CO]].
 \bibitem{Dvali:1998pa}                                                                         
  G.~R.~Dvali and S.~H.~H.~Tye,                                                                
  Phys.\ Lett.\  B {\bf 450}, 72 (1999)                                                        
  [arXiv:hep-ph/9812483].                                                                      
                                                                                               
\bibitem{Alexander:2001ks}                                                                     
  S.~H.~S.~Alexander,                                                                          
  Phys.\ Rev.\  D {\bf 65}, 023507 (2002)                                                      
  [arXiv:hep-th/0105032].                                                                      
                                                                                               
\bibitem{Dvali:2001fw}                                                                         
  G.~R.~Dvali, Q.~Shafi and S.~Solganik,                                                       
  arXiv:hep-th/0105203.                                                                        
\bibitem{Burgess:2001fx}                                                                       
  C.~P.~Burgess, M.~Majumdar, D.~Nolte, F.~Quevedo, G.~Rajesh and R.~J.~Zhang,                 
  JHEP {\bf 0107}, 047 (2001)                                                                  
  [arXiv:hep-th/0105204].                                                                      
                                                                                               
\bibitem{Brodie:2003qv}                                                                        
  J.~H.~Brodie and D.~A.~Easson,                                                               
  JCAP {\bf 0312}, 004 (2003)                                                                  
  [arXiv:hep-th/0301138].                                                                      
                                                                                               
\bibitem{Kachru:2003sx}                                                                        
  S.~Kachru, R.~Kallosh, A.~Linde, J.~M.~Maldacena, L.~P.~McAllister and S.~P.~Trivedi,        
  JCAP {\bf 0310}, 013 (2003)                                                                  
  [arXiv:hep-th/0308055].                                                                      

\bibitem{Silverstein:2003hf}
  E.~Silverstein and D.~Tong,
  Phys.\ Rev.\  D {\bf 70}, 103505 (2004)
  [arXiv:hep-th/0310221].
\bibitem{Alishahiha:2004eh}
  M.~Alishahiha, E.~Silverstein and D.~Tong,
  Phys.\ Rev.\  D {\bf 70}, 123505 (2004)
  [arXiv:hep-th/0404084].
\bibitem{Liddle:1994dx}
  A.~R.~Liddle, P.~Parsons and J.~D.~Barrow,
  Phys.\ Rev.\  D {\bf 50}, 7222 (1994)
  [arXiv:astro-ph/9408015].
\bibitem{Babichev:2007dw} 
  E.~Babichev, V.~Mukhanov and A.~Vikman,
  JHEP {\bf 0802}, 101 (2008)
  [arXiv:0708.0561 [hep-th]].
  
\bibitem{Giddings:2001yu} 
  S.~B.~Giddings, S.~Kachru and J.~Polchinski,
  Phys.\ Rev.\ D {\bf 66}, 106006 (2002)
  [hep-th/0105097].

\bibitem{Douglas:2006es}
  M.~R.~Douglas and S.~Kachru,
  Rev.\ Mod.\ Phys.\  {\bf 79}, 733 (2007)
  [arXiv:hep-th/0610102].
\bibitem{Chen:2005ad}
  X.~Chen,
  JHEP {\bf 0508}, 045 (2005)
  [arXiv:hep-th/0501184].
\bibitem{Garriga:1999vw}
  J.~Garriga and V.~F.~Mukhanov,
  Phys.\ Lett.\  B {\bf 458}, 219 (1999)
  [arXiv:hep-th/9904176].


\bibitem{Agarwal:2008ah}
  N.~Agarwal and R.~Bean,
  Phys.\ Rev.\  D {\bf 79}, 023503 (2009)
  [arXiv:0809.2798 [astro-ph]].


\bibitem{Powell:2008bi}
  B.~A.~Powell, K.~Tzirakis and W.~H.~Kinney,
  JCAP {\bf 0904}, 019 (2009)
  [arXiv:0812.1797 [astro-ph]].

\bibitem{Chen:2006nt}
  X.~Chen, M.~x.~Huang, S.~Kachru and G.~Shiu,
  JCAP {\bf 0701}, 002 (2007)
  [arXiv:hep-th/0605045].
\bibitem{Komatsu:2010fb} 
  E.~Komatsu {\it et al.}  [WMAP Collaboration],
  Astrophys.\ J.\ Suppl.\  {\bf 192}, 18 (2011)
  [arXiv:1001.4538 [astro-ph.CO]].
\bibitem{Baumann:2008aq}
  D.~Baumann {\it et al.}  [CMBPol Study Team Collaboration],
  AIP Conf.\ Proc.\  {\bf 1141}, 10 (2009)
  [arXiv:0811.3919 [astro-ph]].
\bibitem{Chen:2005fe}
  X.~Chen,
  Phys.\ Rev.\  D {\bf 72}, 123518 (2005)
  [arXiv:astro-ph/0507053].

\bibitem{Sefusatti:2009xu}                                                                     
  E.~Sefusatti, M.~Liguori, A.~P.~S.~Yadav, M.~G.~Jackson and E.~Pajer,                        
  JCAP {\bf 0912}, 022 (2009)                                                                  
  [arXiv:0906.0232 [astro-ph.CO]].                                                             

\bibitem{bbo}See: \texttt{http://universe.nasa.gov/program/bbo.html}
\bibitem{Seto:2001qf}                                                                          
  N.~Seto, S.~Kawamura and T.~Nakamura,                                                        
  Phys.\ Rev.\ Lett.\  {\bf 87}, 221103 (2001)                                                 
  [arXiv:astro-ph/0108011].                                                                    
\bibitem{Seto:2005qy}                                                                          
  N.~Seto,                                                                                     
  Phys.\ Rev.\  D {\bf 73}, 063001 (2006)                                                      
  [arXiv:gr-qc/0510067].                                                                       
\bibitem{Kudoh:2005as}                                                                         
  H.~Kudoh, A.~Taruya, T.~Hiramatsu and Y.~Himemoto,                                           
  Phys.\ Rev.\  D {\bf 73}, 064006 (2006)                                                      
  [arXiv:gr-qc/0511145].                                                                       
\bibitem{Hoffman:2000ue}                                                                       
  M.~B.~Hoffman and M.~S.~Turner,                                                              
  Phys.\ Rev.\  D {\bf 64}, 023506 (2001)                                                      
  [arXiv:astro-ph/0006321].                                                                    
\bibitem{Kinney:2002qn}                                                                        
  W.~H.~Kinney,                                                                                
  Phys.\ Rev.\  D {\bf 66}, 083508 (2002)                                                      
  [arXiv:astro-ph/0206032].                                                                    
\bibitem{Easther:2002rw}                                                                       
  R.~Easther and W.~H.~Kinney,                                                                 
  Phys.\ Rev.\  D {\bf 67}, 043511 (2003)                                                      
  [arXiv:astro-ph/0210345].                                                                    

\bibitem{Peiris:2007gz}
  H.~V.~Peiris, D.~Baumann, B.~Friedman and A.~Cooray,
  Phys.\ Rev.\  D {\bf 76}, 103517 (2007)
  [arXiv:0706.1240 [astro-ph]].

\bibitem{Kinney:2007ag}
  W.~H.~Kinney and K.~Tzirakis,
  Phys.\ Rev.\  D {\bf 77}, 103517 (2008)
  [arXiv:0712.2043 [astro-ph]].
\bibitem{Lorenz:2008et}
  L.~Lorenz, J.~Martin and C.~Ringeval,
  Phys.\ Rev.\  D {\bf 78}, 083513 (2008)
  [arXiv:0807.3037 [astro-ph]].
\bibitem{Colombo:2008ta} 
  L.~P.~L.~Colombo, E.~Pierpaoli and J.~R.~Pritchard,
  Mon.\ Not.\ Roy.\ Astron.\ Soc.\  {\bf 398}, 1621 (2009)
  [arXiv:0811.2622 [astro-ph]].
\bibitem{Bond:2004rt}
  J.~R.~Bond, C.~R.~Contaldi, A.~M.~Lewis and D.~Pogosyan,
  Int.\ J.\ Theor.\ Phys.\  {\bf 43}, 599 (2004)
  [arXiv:astro-ph/0406195].
\bibitem{Dodelson:1997hr}                                                                      
  S.~Dodelson, W.~H.~Kinney and E.~W.~Kolb,                                                    
  Phys.\ Rev.\  D {\bf 56}, 3207 (1997)                                                        
  [arXiv:astro-ph/9702166].                                                                    
\bibitem{Kinney:2003uw}                                                                        
  W.~H.~Kinney, E.~W.~Kolb, A.~Melchiorri and A.~Riotto,                                       
  Phys.\ Rev.\  D {\bf 69}, 103516 (2004)                                                      
  [arXiv:hep-ph/0305130].                                                                      
\bibitem{Zhao:2009rt}                                                                          
  W.~Zhao and W.~Zhang,                                                                        
  Phys.\ Lett.\  B {\bf 677}, 16 (2009)                                                        
  [arXiv:0907.1453 [astro-ph.CO]].                                                             
\bibitem{Zhao:2009mj}
  W.~Zhao and D.~Baskaran,
  Phys.\ Rev.\  D {\bf 79}, 083003 (2009)
  [arXiv:0902.1851 [astro-ph.CO]].
  
\bibitem{Dent:2012ne} 
  J.~B.~Dent, D.~A.~Easson and H.~Tashiro,
  Phys.\ Rev.\ D {\bf 86}, 023514 (2012)
  [arXiv:1202.6066 [astro-ph.CO]].
  
\bibitem{Cooray:2006km} 
  A.~Cooray,
  Phys.\ Rev.\ Lett.\  {\bf 97}, 261301 (2006)
  [astro-ph/0610257].
  
\bibitem{Adshead:2010mc} 
  P.~Adshead, R.~Easther, J.~Pritchard and A.~Loeb,
  JCAP {\bf 1102}, 021 (2011)
  [arXiv:1007.3748 [astro-ph.CO]].

\end{thebibliography}
\end{document}